\documentclass{aa}
\usepackage[varg]{txfonts}
\usepackage{graphicx}
\usepackage{subfigure}
\usepackage{rotating}
\usepackage{xcolor}
\usepackage{natbib}
\usepackage{hyperref}
\usepackage{float}
\usepackage{wrapfig,lipsum}
\usepackage{inputenc}
\usepackage{multirow}
\usepackage{siunitx}
\DeclareUnicodeCharacter{2212}{-}
\usepackage{hyperref}
\usepackage[flushleft]{threeparttable} 
\usepackage{ulem}

\bibpunct{(}{)}{;}{a}{}{,} 

\definecolor{darkspringgreen}{rgb}{0.09, 0.45, 0.27}

\begin{document} 

\title{Two interacting galaxies hiding as one, revealed by MaNGA}
\author{Barbara Mazzilli Ciraulo
          \inst{1}
          \and
          Anne-Laure Melchior\inst{1}
          \and
          Daniel Maschmann\inst{1}
          \and
          Ivan Yu. Katkov\inst{2,3,4}
          \and
          Ana\"elle Halle\inst{1}
          \and
          Fran\c coise Combes\inst{1}\fnmsep\inst{5}
          \and
          Joseph D. Gelfand\inst{2,3,6}
          \and
          Aisha Al Yazeedi\inst{2,3}
          }
          
\institute{Observatoire de Paris, PSL university, Sorbonne Universit\'e, CNRS, LERMA, F-75014, Paris, France \\
              \email{Barbara.Mazzilli-Ciraulo@observatoiredeparis.psl.eu}         \and
         New York University Abu Dhabi, Saadiyat Island, PO Box 129188, Abu Dhabi, UAE
         \and
         Center for Astro, Particle, and Planetary Physics, NYU Abu Dhabi, PO Box 129188, Abu Dhabi, UAE
         \and
         Sternberg Astronomical Institute, M.V. Lomonosov Moscow State University, 13 Universitetsky prospect, Moscow, 119991, Russia
         \and
         Coll \`{e}ge de France, 11 Place Marcelin Berthelot, 75005 Paris, France
         \and
         Center for Cosmology and Particle Physics, New York University, 726 Broadway, New York, NY 10003, US\\
}

\abstract{
Given their prominent role in galaxy evolution, it is of paramount importance to unveil galaxy interactions and merger events and to investigate the underlying mechanisms.
The use of high-resolution data makes it easier to identify merging systems, but it can still be challenging when the morphology does not show any clear galaxy pair or gas bridge. Characterising the origin of puzzling kinematic features can help reveal complicated systems.
These counter-rotating objects both lie on the star-forming main sequence but display perturbed stellar velocity dispersions. The main galaxy presents off-centred star formation as well as off-centred high-metallicity regions, supporting the scenario of recent starbursts, while the secondary galaxy hosts a central starburst that coincides with an extended radio emission, in excess with respect to star formation expectations. Stellar mass as well as dynamical mass estimates agree towards a mass ratio within the visible radius of 9:1 for these interacting galaxies.
We suggest that we are observing a pre-coalescence stage of a merger. The primary galaxy accreted gas through a past first pericentre passage about 1 Gyr ago and more recently from the secondary gas-rich galaxy, which exhibits an underlying active galactic nucleus.
Our results demonstrate how a galaxy can hide another one and the relevance of a multi-component approach for studying ambiguous systems.
We anticipate that our method will be efficient at unveiling the mechanisms taking place in a sub-sample of galaxies observed by the Mapping Nearby Galaxies at Apache Point Observatory (MaNGA) survey, all of which exhibit kinematic features of a puzzling origin in their gas emission lines.
}
 
\keywords{ Galaxy: evolution --
                Galaxy: kinematics and dynamics--
                Galaxies: interactions --
                Techniques: spectroscopic --
                Methods: data analysis
               }
\maketitle

\section{Introduction}
The importance of galaxy interactions and merger events is nowadays beyond dispute as hierarchical galaxy evolution is widely adopted and multiple observational as well as theoretical works have been carried out to investigate the various mechanisms related to the evolution of interacting galaxies \citep[e.g.][]{2015ApJ...813...23V,2016ARA&A..54..597C}.
The improvement in data resolution has made it possible to highlight various observational signatures of mergers, such as colour change \citep{2012A&A...539A..46A}, morphological disruption \citep{2010MNRAS.407.1514E,2013MNRAS.429.1051C}, star formation enhancement \citep{2013MNRAS.433L..59P}, and merger-induced nuclear activity at low redshift for optical and mid-IR active galactic nuclei \citep[AGNs;][]{2019MNRAS.487.2491E}.
The presence of double nuclei \citep[such as in NGC 3526;][]{2014ApJ...797...90S} is a clear signature of mergers as the existence of tidal tails \citep{2014MNRAS.438.1784M} and, alternatively, spectral features, such as double-peaked profiles in molecular gas emission lines, have also been interpreted as merger signatures \citep[e.g.][]{2005MNRAS.359.1165G,2005A&A...440L..45W}. While the timescales of the merging are well monitored statistically through numerical simulations \citep[e.g.][]{2015ARA&A..53...51S,2019ApJ...872...76N} and for individual objects \citep[e.g.][]{2018MNRAS.475.3934L}, dating the epochs of these observed processes remains a challenge. 
Lastly, some important questions remain unanswered regarding the possible impact of AGNs on the merging stage \citep[e.g.][]{2007MNRAS.375.1017A, 2020MNRAS.494.5713M,2020ApJ...904..107S,2021arXiv210101729S} and whether the quenching is related to the merging \citep[e.g.][]{2018MNRAS.478.3447E,2020MNRAS.493.3716H,2021arXiv210102564D}. 


\begin{table*}[h]
    \caption{Information about MaNGA ID 1-114955.}
    \label{tab:merger_info}
    \vspace{-0.6cm}
    \begin{center}
    \begin{tabular}{cccccccc}
    \hline \hline
        r.a. & dec. & redshift & morphology & inclination & $M_{\star}$ & SFR$_{\rm SED}$ & SFR\\
        (J2000) & (J2000) & & & ($^{\circ}$) & ($M_{\odot}$) & ($M_{\odot} \, \mathrm{yr^{-1}}$) & (${M_{\odot}  \, \mathrm{yr^{-1}}}$) \\
    \hline
        22h10m24.5s & +11d42m47s & 0.09228 & interacting & 44 & $1.59\times10^{11}$ & 13.3 & 10.6\\
    \hline
    \end{tabular}
    \end{center}
    \vspace{0.1cm}
    {\small
    {{\bf Notes:} The coordinates and redshift come from the NASA-Sloan Atlas catalogue, while the morphology is determined by visual inspection (this object is classified as uncertain in Galaxy Zoo). The inclination is derived using the measurements from \citet{2018MNRAS.476.3661D}. The stellar mass and the star formation rate (SFR) value are  derived from SED fitting based on optical and UV spectra \citep{2016ApJS..227....2S}. The last column gives the SFR estimate outlined in \citet{2004MNRAS.351.1151B}.}}
\end{table*}
\begin{figure}[h]
    \centering
    \includegraphics[width=.5\textwidth]{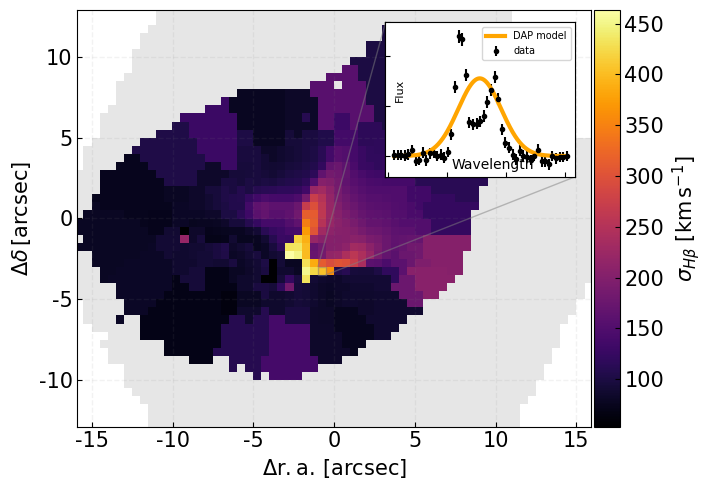}
    \vspace{-0.5cm}
    \caption{Velocity dispersion of the $\mathrm{H \beta}$ emission line derived by the MaNGA DAP. The MaNGA field of view is represented as a grey footprint. The included panel shows a portion of the continuum-subtracted spectrum from one spaxel where the model (in green) does not manage to fit the spectral features.}
    \label{fig:sigma_dap}
\end{figure}

In this article we present a multi-wavelength analysis of a galaxy merger (J221024.49+114247.0) at $z=0.09$, identified through the detection of double-peaked emission-line profiles in the central spectrum of the Sloan Digital Sky Survey \citep[SDSS;][]{2002AJ....124.1810S}. Two gas and stellar counter-rotating components have been identified and analysed. In Sect. \ref{sect:ana} we discuss how we extract properties from the Mapping Nearby Galaxies at Apache Point Observatory  \citep[MaNGA;][]{2015ApJ...798....7B} data cube to identify the two components. In Sect. \ref{sect:other} we discuss the molecular gas data, as well as other archives (Very Large Array) and properties used from value-added catalogues. In Sect. \ref{sect:results} we analyse the properties of this merger. In Sect. \ref{sect:disc} we discuss our results.
We adopt throughout this work a $\Lambda$ cold dark matter (CDM) cosmology with parameters $\Omega_\Lambda$ = 0.7, $\Omega_M$ = 0.3, and $H_0 = 70\, \mathrm{km}\,\mathrm{s}^{-1}\,\mathrm{Mpc}^{-1}$.
 
\begin{figure*}[h!]
\begin{minipage}{\textwidth}
\centering
    \includegraphics[width=\textwidth]{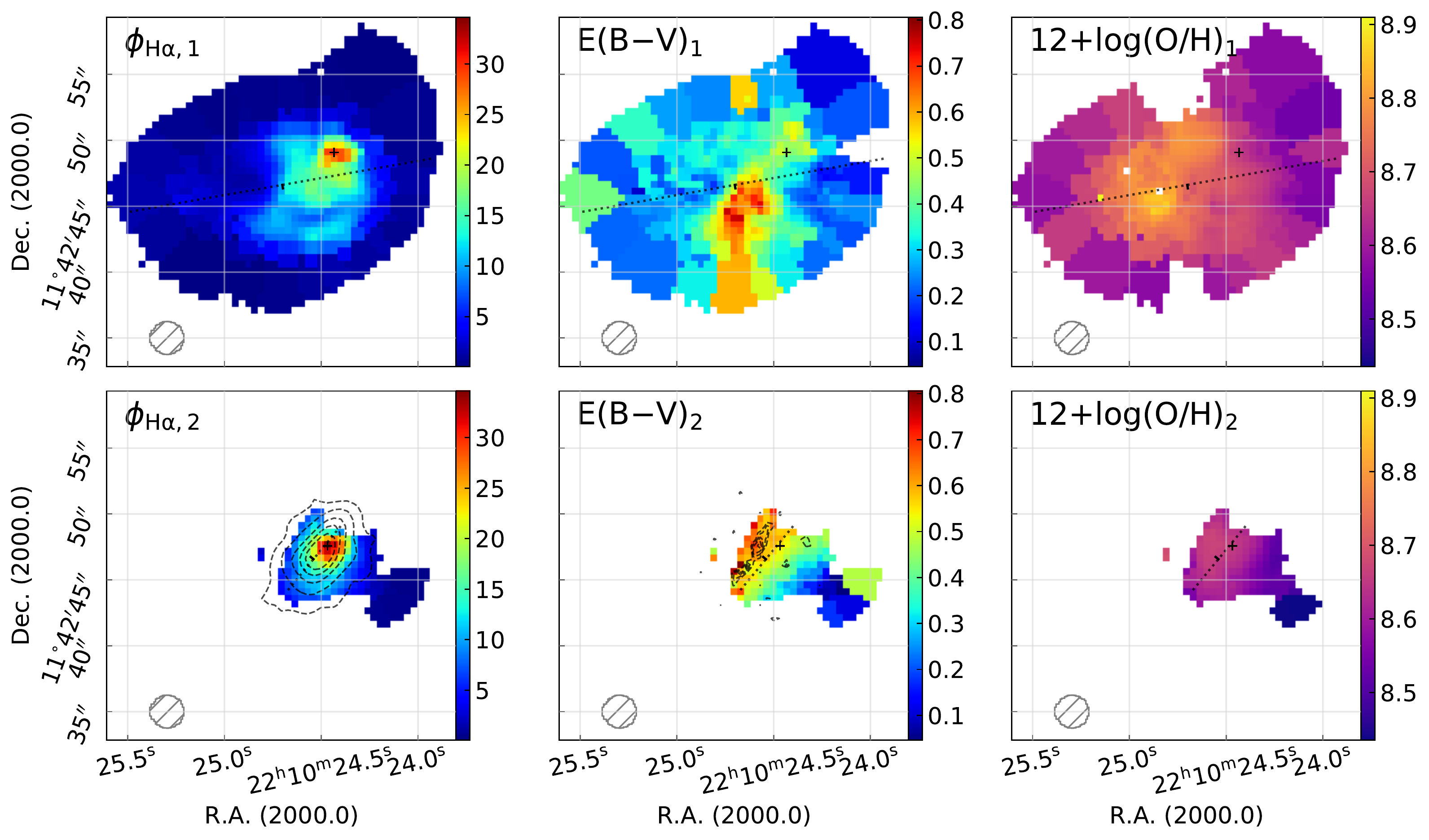}
    \vspace{-0.5cm}
    \caption{Gas properties derived from our multi-component approach. The first and second rows show the maps for the first and second components detected in MaNGA data, respectively. The first column presents the $\mathrm{H \alpha}$ extinction-corrected flux (in ${\rm erg\,s^{-1}\,\AA^{-1}\,cm^{-2}}$ per spaxel), the second column displays the  extinction computed from the Balmer decrement, and the last column shows the oxygen gas-phase abundance derived using the O$_3$N$_2$ calibrator. 
    The black crosses indicate the position of the extinction-corrected $\mathrm{H \alpha}$ flux peak for the represented component. The MaNGA PSF is displayed as a hatched grey circle in the bottom-left corner of the panels.}
    \label{fig:gas_2comp_properties}
\end{minipage}
\\
\begin{minipage}{\textwidth}
    \centering
    \includegraphics[width=\textwidth]{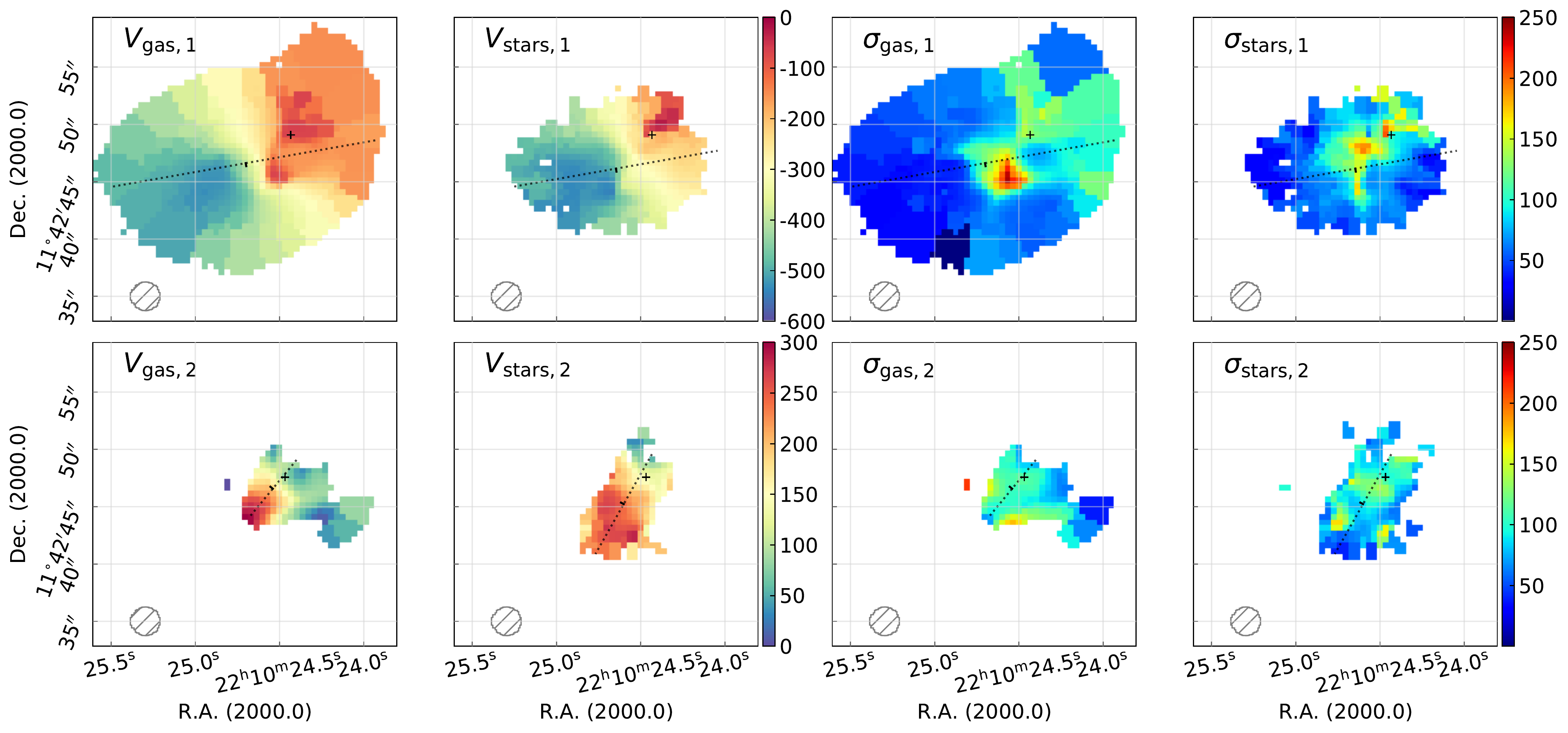}
    \vspace{-0.5cm}
    \caption{Gas and star kinematics. The first row shows the maps for the first component, and the second row the second component detected in MaNGA data. The first and second columns display the respective ionised gas and stellar velocity fields (in km\,s$^{-1}$). The dotted black lines refer to the computed position angles. The third and fourth columns show the respective velocity dispersion fields (in km\, s$^{-1}$) for the gas and the stars. As in Fig. 2, the black crosses indicate the position of the extinction-corrected $\mathrm{H \alpha}$ flux peak for the represented component.}
    \label{fig:gas_stars_kinematics}
    \end{minipage}
\end{figure*}
\vspace{-0.5cm}

\section{MaNGA data analysis}
\label{sect:ana}
The galaxy J221024.49+114247.0 is one optical merger identified in a sub-sample (Mazzilli Ciraulo et al, in prep.) obtained by cross-matching the summary file from the MaNGA  Data Reduction Pipeline \citep[DRP; SDSS Data Release DR14, ][]{2018ApJS..235...42A} and a double-peak galaxy catalogue produced by \citet{2020A&A...641A.171M}, relying on the Reference Catalogue of Spectral Energy Distributions (RCSED) from \citet{2017ApJS..228...14C}. The basic properties of this source, provided by different value-added catalogues, are given in Table \ref{tab:merger_info}. The Legacy survey \citep{2019AJ....157..168D} reveals a disrupted object, and the integrated spectrum within the 3 arcsec SDSS fibre shows double-peaked emission lines. 
The effective point spread function (PSF) of the data is about 2.5 arcsec and corresponds to a spatial resolution of 4.3\,kpc at the redshift of the source ($z=0.09228$), and the data are gridded on $0.5\, \times\, 0.5^{\prime\prime}$ spaxels.

We produced 2D maps based on the derived properties from the MaNGA Data Analysis Pipeline \citep[DAP; SDSS-IV,][]{ 2019AJ....158..231W}. The main emission lines peak in a compact central region. The velocity field does not display any standard rotating motion and appears to be strongly irregular, with values covering a very wide range ($ -525 \, \mathrm{km \, s^{-1}} \, < V_{\rm gas} < \, 445 \, \mathrm{km \, s^{-1}}$). As displayed in Fig. \ref{fig:sigma_dap}, the velocity dispersion is also remarkable, with spaxels exhibiting values greater than ${\rm 400\,km\,s^{-1}}$ and not centred. The double-peak features of these spaxels are not well fitted by the single component fit, and some of them are masked during the data processing. In order to further investigate this complex system, we developed an optimised multi-component analysis.

\subsection{Gas kinematics derived from the multi-component approach}
\label{sect:evolution_fit}
We developed a fitting procedure, adjusting two components for the selected emission lines. We ran and applied it to the Voronoi binned continuum-subtracted spectra and fitted $\mathrm{H \beta}$, [OIII] doublet, $\mathrm{H \alpha}$, and [NII] doublet lines. As in the MaNGA DAP, we assigned the same velocity and velocity dispersion to all these lines.
We were able to highlight the presence of two components along the line of sight.
We first selected reliable spaxels, relying on a signal-to-noise ratio (S/N) threshold of 3. When the S/N criterion is not met by the two fitted components, we kept the single-Gaussian results and associated them with the main galaxy.
For the secondary component, we based our spaxel selection on an F-test \citep{mendenhall2011second} to ensure that the double-Gaussian fit was better than the single Gaussian one, that a velocity difference between both peaks was significantly greater than the MaNGA data spectral resolution, and that there was an amplitude ratio between both peaks to guarantee that one of them was not completely suppressed.

We display the derived properties for each of them on the 2D maps shown in Figs. \ref{fig:gas_2comp_properties} and \ref{fig:gas_stars_kinematics}. Figure \ref{fig:gas_2comp_properties} displays the extinction-corrected $\mathrm{H \alpha}$ line fluxes, the extinction derived from the Balmer decrement, and the gas metallicity using the O$_3$N$_2$ calibrator, while Fig. \ref{fig:gas_stars_kinematics} exhibits the gas kinematics of the two components with their velocity and velocity dispersion.
The spatially resolved maps confirm the contribution of two separate components: The first is related to a galaxy with a fairly regular velocity field, whereas the second is detected in a smaller region of the field of view, shows a slightly stronger $\mathrm{H \alpha}$ line flux, and shows a velocity field counter-rotating with respect to the main object. 


\subsection{Multi-component analysis of the stellar kinematics}
\label{sect:nonparam_losvd}
To explore whether the double-peaked structure in the emission lines has associated counterparts in the stellar kinematics, we applied an analysis workflow used for the study of galaxies that host counter-rotating stellar discs \citep{2013ApJ...769..105K_ic719, 2016MNRAS.461.2068K_n448}. Firstly, we fitted spectra using a full spectral fitting technique, \textsc{NBursts} \citep{2007IAUS..241..175C,2007MNRAS.376.1033C}, in the basic mode, where a spectrum is approximated by a single simple stellar population (SSP) model. Secondly, we recovered a non-parametric stellar line-of-sight velocity distribution (LOSVD) using an un-broadened stellar population model template from the first step \citep[see method details in][]{2020MNRAS.493.5464K_n7572}. The non-parametrical stellar LOSVD reveals a clearly separated double-peaked structure, on which we adjusted a double-Gaussian function to derive kinematics parameters of individual components.
We hence spotlight two stellar population components: One is regular, and the second is less prominent but is detected in the region where we see a second gas component as well. We used quality filters based on the uncertainties of the parameters and a sufficient relative flux for the second component to show relevant 2D maps.

The second and fourth columns of Fig. \ref{fig:gas_stars_kinematics} display the stellar velocity and stellar velocity dispersion fields, respectively, for each component.
Finally, we re-fitted spectra with \textsc{NBursts}, using a two-component approach in which both components have individual and independent kinematics. We considered velocity and dispersion estimates from the previous step as initial guesses. This analysis suggests that both components have similar but slightly different (SSP-equivalent) averaged stellar ages of $ T_1 \approx 1.70 \, \mathrm{Gyr}$ and $ T_2 \approx 1.42 \, \mathrm{Gyr}$ and metallicities of ${\rm [Fe/H]_1\approx -0.24\,dex}$ and ${\rm [Fe/H]_2\approx-0.82\,dex}$.

\section{Analysis of other data}
\label{sect:other}
\subsection{Molecular gas content}
\begin{figure}[h]
    \centering
    \includegraphics[width=0.48\textwidth]{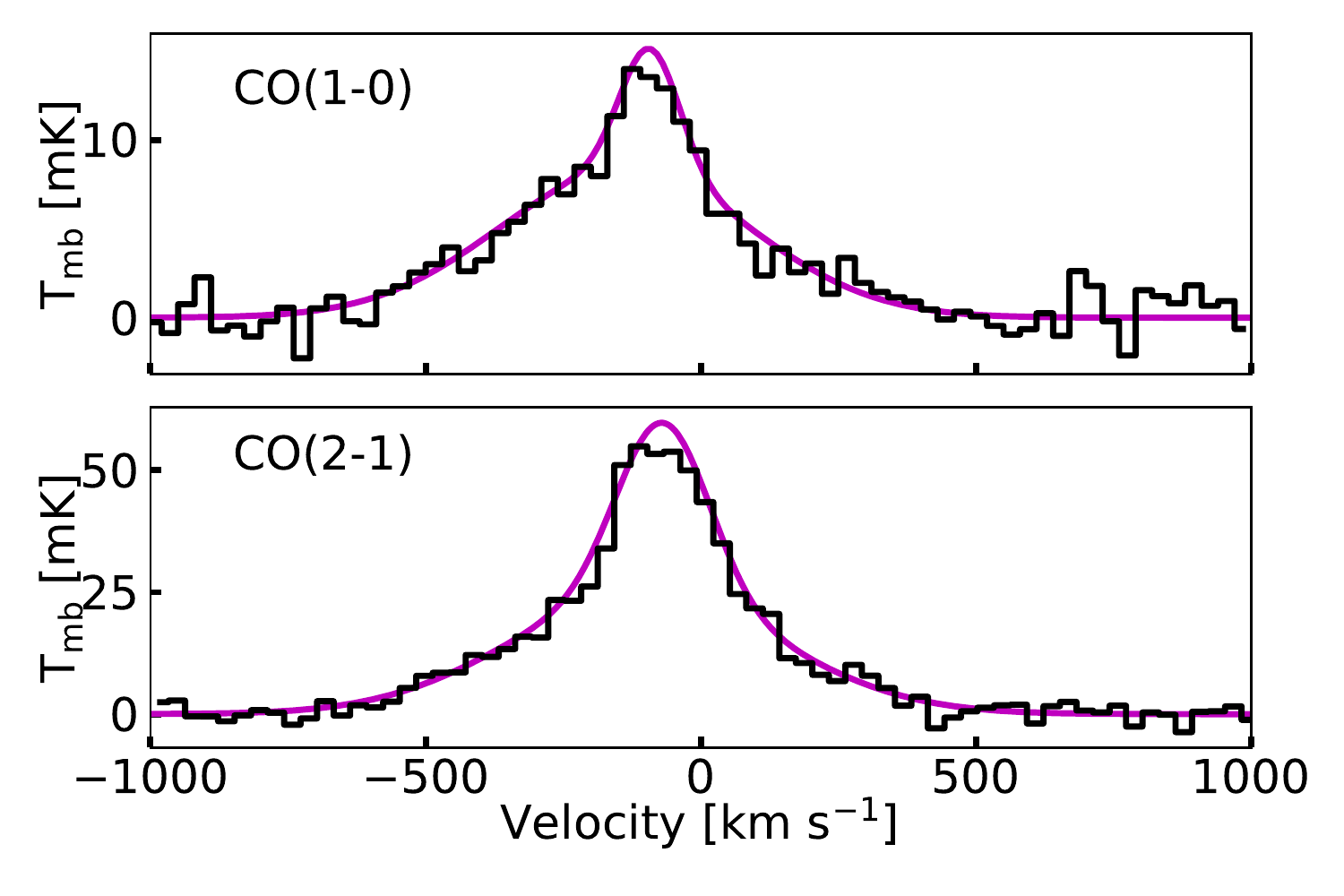}
    \caption{
    Molecular gas. We show the CO(1-0) and CO (2-1) lines observed with the IRAM 30m telescope in units of main beam temperature, T$_{\rm mb}$.     
    }
    \label{fig:double_fitted_co_spectra}
\end{figure}
IRAM 30m telescope observations were carried out in December 2019 with the Eight MIxer Receiver (EMIR) and the symmetrical wobbler switching mode, in which the secondary mirror switches up to a maximum of $\pm$120\,arcsec in azimuth. We operated in dual-sideband mode, and these sidebands were connected to two spectroscopic back ends, WILMA and FTS.
J221024.49+114247.0 was integrated for 2.0 hours in the E090 band and 2.8 hours in the E230 band, enabling the detection of CO(1-0) emission in the former and CO(2-1) emission in the latter, as displayed in Fig. \ref{fig:double_fitted_co_spectra}.
We fitted linear baselines and subtracted them from the spectra. The lines exhibit wide tails and are better adjusted with two Gaussian functions of different widths (see Fig. \ref{fig:double_fitted_co_spectra} and the parameters in Table \ref{tab:double_fitted_co_spectra}) than with a single Gaussian function fit.  Firstly, we calculated the CO line luminosity, in $\textrm{K}\ \textrm{km}\ \textrm{s}^{-1}\ \textrm{pc}^2$, as defined in \cite{1997ApJ...478..144S}:
\begin{equation}
   L_{\rm CO} = 3.25\ \times 10^7\,S_{\rm CO}\, \Delta V\, \nu_{\rm obs}^{-2}\, D_L^{2}\,(1+z)^{-3}
\end{equation}
where $S_{\rm CO}=5\,I_{\rm CO}$ is the integrated flux in Jy\,km\,s$^{-1}$, $I_{\rm CO}$ is the line area given in Table \ref{tab:double_fitted_co_spectra} in K\,km\,s$^{-1}$ (we considered the sum of both $I_{\rm CO}$ in order to integrate the whole flux), $\nu_{\rm obs}$ is the line frequency in GHz, and $D_{L}=422 \, \mathrm{Mpc}$ is the luminosity distance to the galaxy. Secondly, we derived the mass of molecular gas using $M_{\rm H_2} = \alpha\ L_{\rm CO}$, where $\alpha$ is the CO-to-H$_2$ conversion factor, estimated following \citet{2019A&A...622A.105F}.
We hence derived a CO-to-H$_2$ conversion factor of $\alpha = 4 \, M_{\odot}/( \mathrm{ K\,km\,s^{-1}\,pc^{2}})$ and finally found a total molecular gas mass of $ M_{\rm H_2} = 3.29 \times 10^{10} \, M_{\odot}$, which corresponds to $21\,\%$ of the total stellar mass.
Table \ref{tab:double_fitted_co_spectra} gives the properties of the reduced CO spectra.
\begin{table}[h]
    \centering
    \caption{Fit parameters for CO(1-0) and CO(2-1) spectra.}
    \begin{tabular}{c|ccccc}
    \hline
    & $V_0$ & $T_{\rm mb}$ & $\sigma_{\rm noise}$ & FWHM$_v$ & $I_{\rm CO}$   \\
    & (km\,s$^{-1}$) & (mK) & (mK) & (km\,s$^{-1}$) & (K\,km\,s$^{-1}$)\\ \hline
\multirow{2}{*}{CO(1-0)} & -139$\pm$14 & 8 & 0.7 & 537 & 4.46$\pm$0.29 \\                         & -92$\pm$8  & 7 & 0.7 & 127 & 0.89$\pm$0.24 \\ \hline
\multirow{2}{*}{CO(2-1)} & -102$\pm$9 & 23  & 1.5 & 571 & 13.83$\pm$0.75 \\
                         &  -69$\pm$3  & 34 & 1.5 & 196 & 7.01$\pm$0.72  \\ \hline
    \end{tabular}
    \label{tab:double_fitted_co_spectra}
\end{table}

\subsection{Faint structures detected in the $r$-band MegaPipe/CFHT image}
\label{sect:cfht}
\begin{figure}[h]
    \centering
    \includegraphics[width=.48\textwidth]{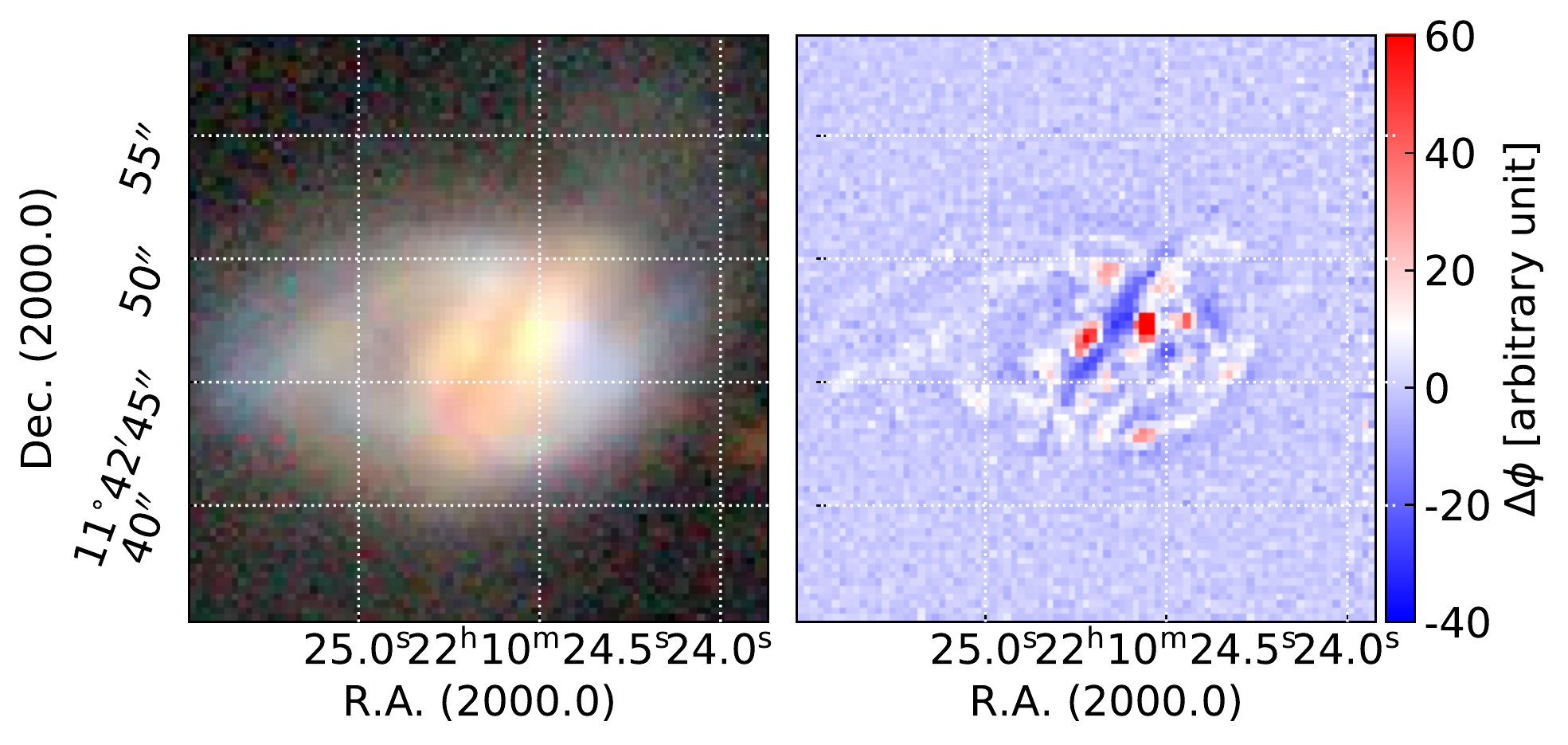}
    \caption{Morphology. \textbf{Left:} Composite image ($g$, $r$, $z$ bands) from the Legacy survey \citep{2019AJ....157..168D}. \textbf{Right:} Difference between the MegaCam r-band filter image and its smoothed version.}
    \label{fig:rgb_legacy_rdiff_megapipe}
\end{figure}

We also used the publicly available $r$-band stacked image from MegaCam, a wide-field imager mounted on the Canada-France-Hawaii Telescope (CFHT), with very precise astrometric and photometric calibrations, respectively accurate to within 0.15\,$^{\prime\prime}$ and 0.03\,mag.
By smoothing the image from the MegaCam Image Stacking Pipeline \citep[MegaPipe;][]{2008PASP..120..212G} using a Gaussian kernel and then subtracting the result from the original image, we spotlight the presence of an obscured dust lane across the galaxy (see the right panel of Fig. \ref{fig:rgb_legacy_rdiff_megapipe}) and some irregular arms of stars around the core of the system.\\

\subsection{Radio continuum VLA archives}
\label{sect:vla}
J221024.49+114247.0 was observed with the Very Large Array (VLA) on 2016 September 30 in A configuration in both frequencies 1.5\,GHz and 2.5\,GHz (which respectively correspond to the $L$ and the $S$ band), with a $1.3\times1.3^{\prime\prime}$ resolution for the former and a $0.65\times0.65^{\prime\prime}$ resolution for the latter (P.I.: J. D. Gelfand). The corresponding fluxes are respectively 8.2\,mJy and 3.1\,mJy. In parallel, we collected the following integrated fluxes from archival data: 7.4$\pm$0.3\,mJy at 1.4\,GHz from the Faint Images of the Radio Sky at Twenty-Centimeters (FIRST) survey \citep{1994ASPC...61..165B, 2015ApJ...801...26H} and 3.1$\pm$0.5\,mJy at 3\,GHz from the Very Large Array Sky Survey \citep[VLASS;][]{2020PASP..132c5001L}. We derived a synchrotron spectral index of -1.3 and a rest-frame 1.4\,GHz luminosity of $1.3 \times 10^{23}$\ W\,Hz$^{-1}$ \citep{2019ApJ...872..148C}. As displayed in the bottom-left panel of Fig. \ref{fig:gas_2comp_properties}, the 1.5\,GHz peak coincides with the second $\mathrm{H \alpha}$ component. 

\section{Results: Properties of the merger}
\label{sect:results}
\subsection{Kinematics and geometry}
\label{ssect:kinegeo}
\subsubsection{Two stellar and gaseous discs}
We highlight the superposition of two components along the line of sight in MaNGA data and interpret these spectral features as the two progenitors of an on-going merger event. This interaction involves one galaxy whose gas kinematics appears fairly regular and a second galaxy that is counter-rotating with respect to the first one. As displayed in Fig. \ref{fig:gas_stars_kinematics}, the stellar velocity motions for both components are consistent with the velocity fields of the gas. The position angles ($PA$) of the gas and stellar velocity fields, computed using the algorithms described in \citet{2006MNRAS.366..787K} \citep[see \texttt{FIT\_KINEMATIC\_PA} in][]{2017MNRAS.466..798C}, are similar: $ PA_{\rm  gas,1}=280.0\pm0.5^{\circ}$ and $PA_{\rm stars,1}=280.0\pm0.5^{\circ}$ for the first component, and 
$ PA_{\rm gas,2}=140.5\pm2.5^{\circ}$ and $PA_{\rm stars,2}=150.5\pm2.0^{\circ}$ for the second component.
Using the axes found for the gas kinematics, we estimated lengths of the major and minor axes of, respectively, 41\,kpc and 27 kpc for the main galaxy, and of 13\,kpc and 9\,kpc for the second galaxy. 

For the main galaxy, we derived an inclination of $i_1=49^\circ$, assuming a negligible disc thickness \citep[e.g.][]{1926ApJ....64..321H} of up to 53$^{\circ}$ with the correction proposed by \citet{1980ApJ...237..655A}. For the second component, we derived an inclination $i_2=46^\circ$ using the basic formula and $i_2=50^\circ$ when applying the correction for disc thickness.

\subsubsection{Dust lane}
\begin{figure}[h]
    \centering
    \includegraphics[width=0.48\textwidth]{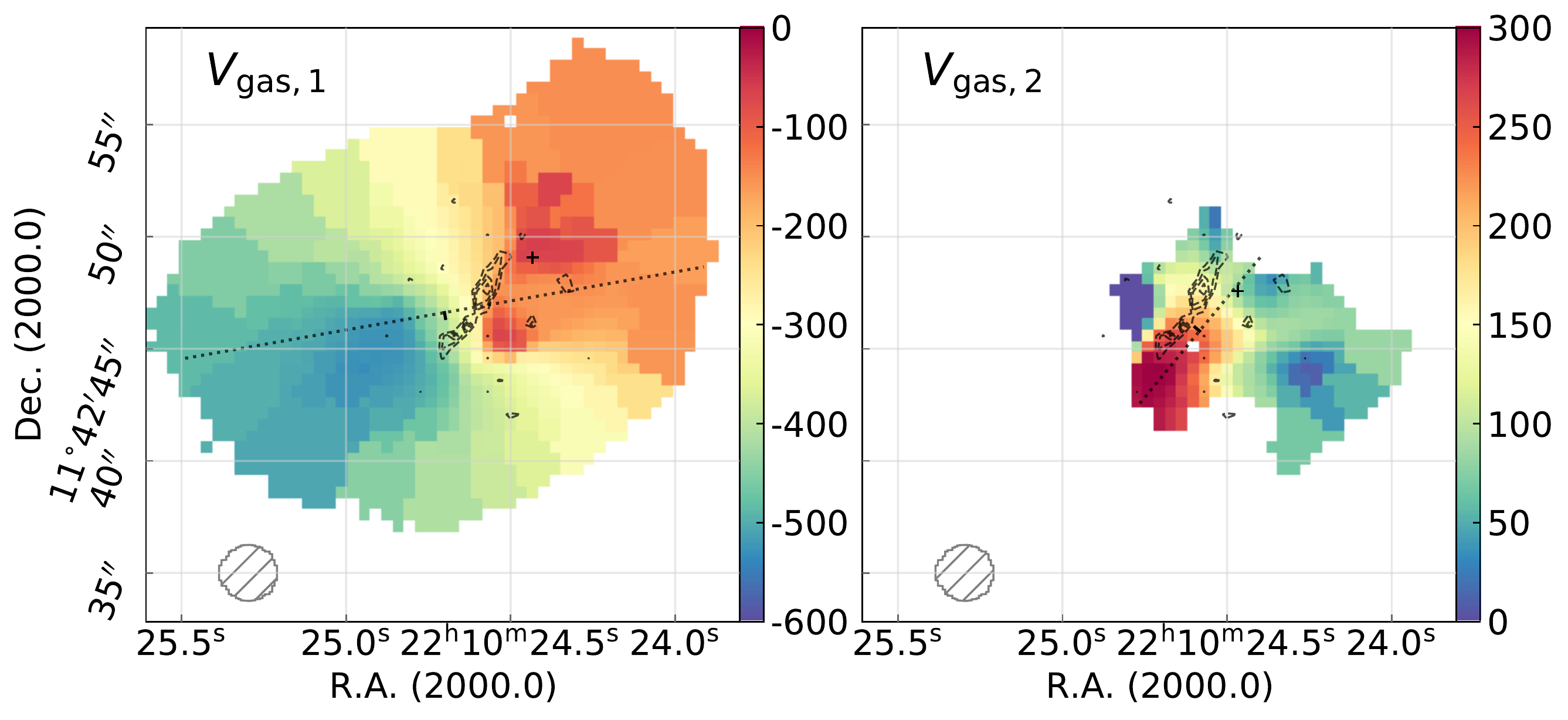}
    \includegraphics[width=0.48\textwidth]{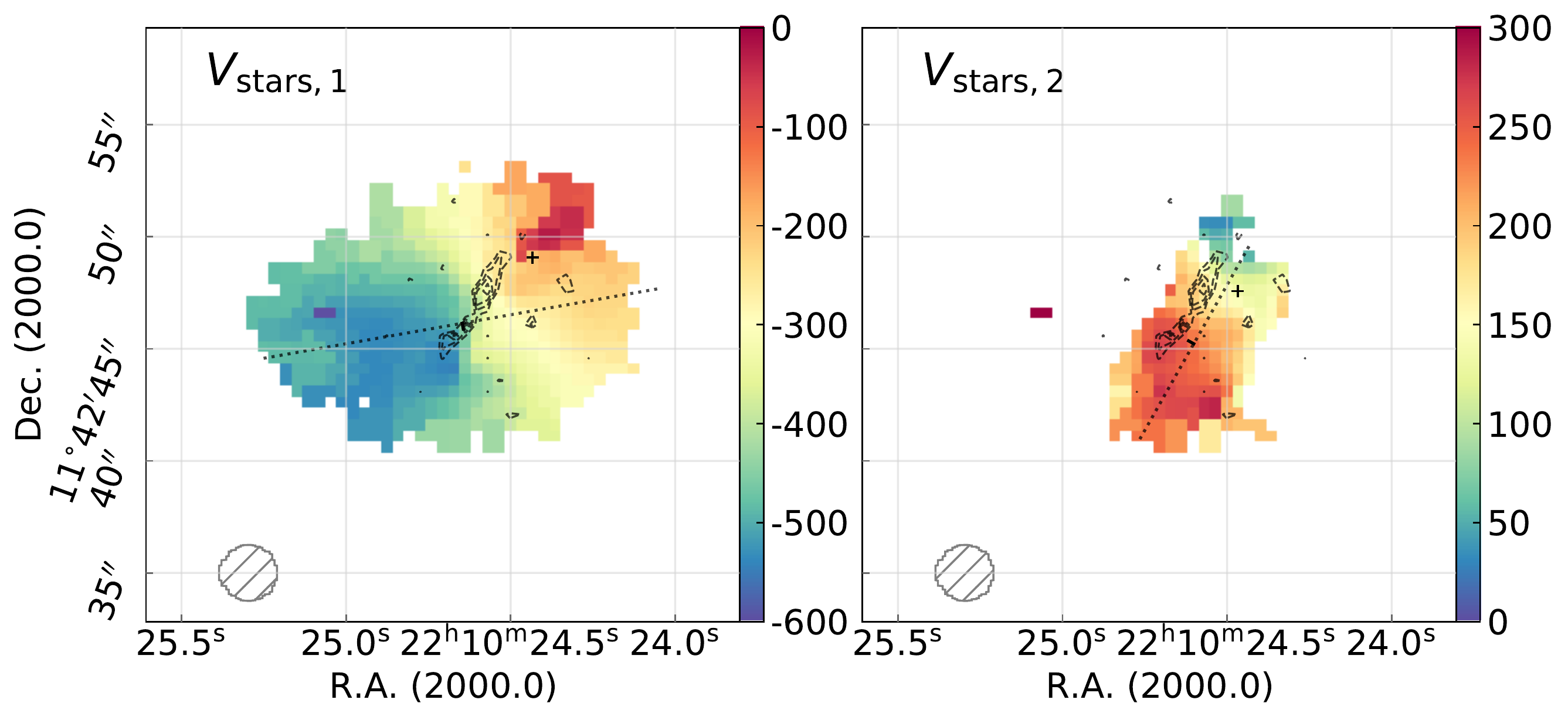}
      \includegraphics[width=0.48\textwidth]{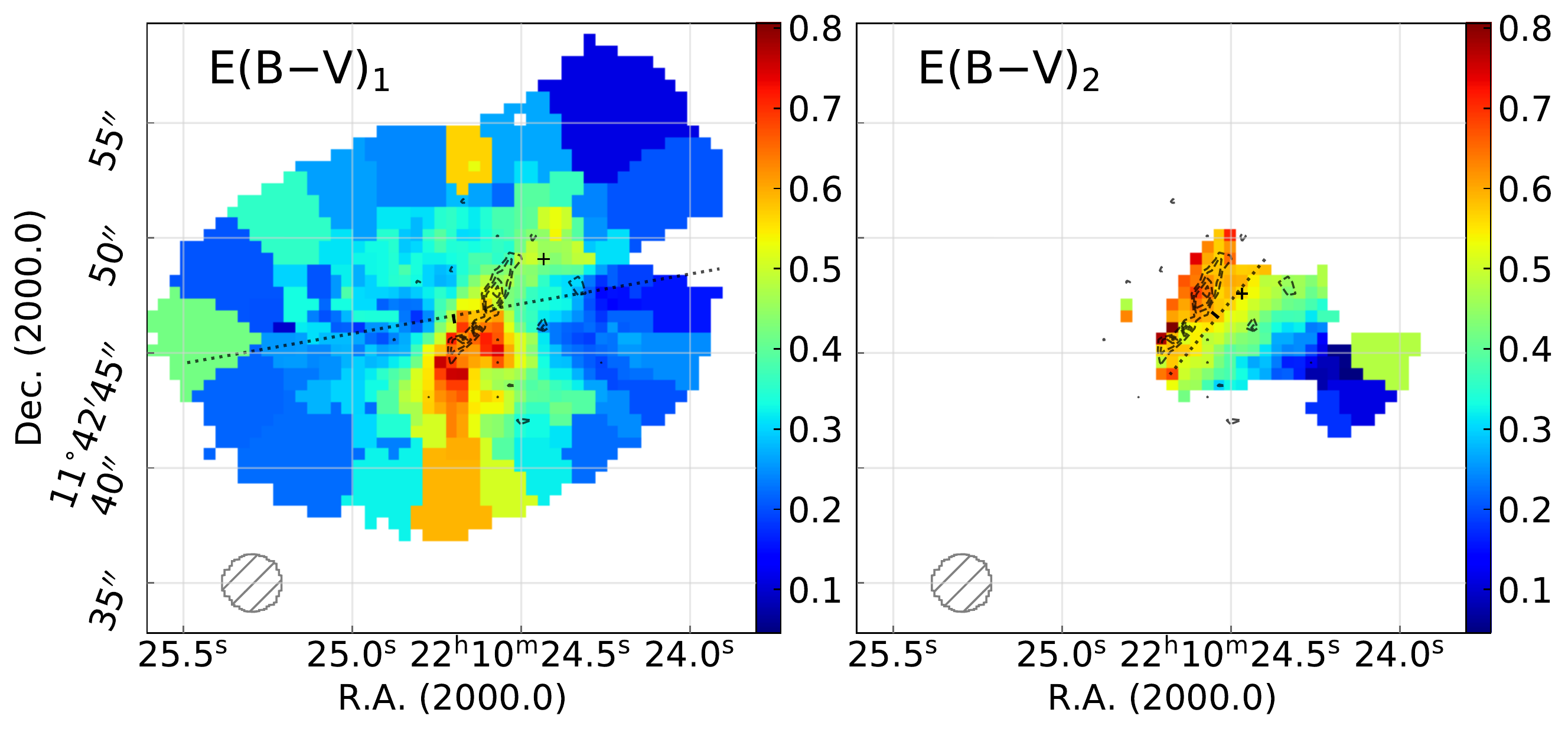}
    \caption{Superimposition of the dust lane on the gas velocity maps (top), the stellar velocity maps (middle), and the extinction maps (bottom) derived from Balmer decrement computation.
    }
    \label{fig:dust-lane}
\end{figure}

The dust lane displayed in the right panel of Fig. \ref{fig:rgb_legacy_rdiff_megapipe} extends from the south-east to the north-west over approximately 11\ kpc. We estimated a position angle  ($PA_{\rm dl}=147^{\circ}$)  consistent with the position angle derived from the velocity field of the second MaNGA component. As displayed in Fig. \ref{fig:dust-lane}, this dust lane is compatible with the extinction computed with the Balmer decrement for the second component, with the same inclination, while it is somehow detected in the extinction measured for the first component, supporting the view that the second component lies in front of the system.  When compared with the stellar and gaseous component (middle and top panels), it exhibits the same position angle as the velocities but shifted to the north, and hence compatible with an inclination of about 50$^\circ$.


\subsection{Star formation activity}
\begin{figure*}[h]
\vspace{-0.3cm}
   \includegraphics[width=.9\textwidth]{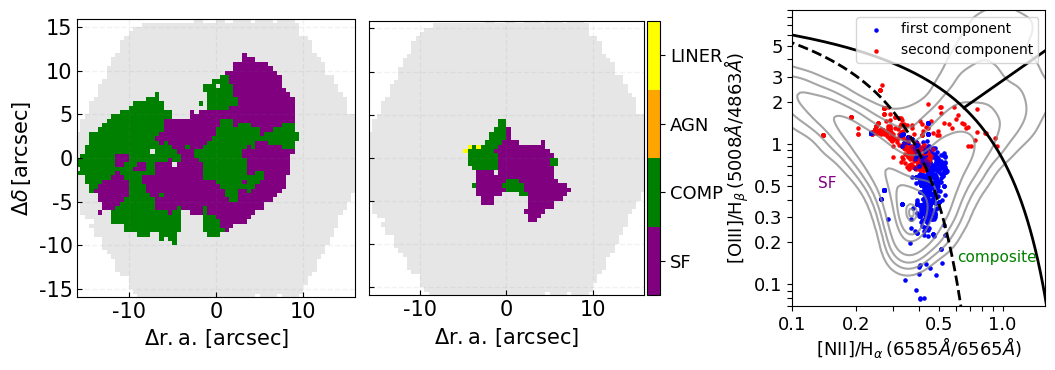}\\
 \vspace{-0.5cm}
   \caption{Ionisation properties of the gas for the two components of J221024.49+114247.0 \textbf{Left  and middle panels:} BPT \citep{1981PASP...93....5B} maps for both fitted components. \textbf{Right panel:} BPT diagram showing the position of the spaxels. The S/N threshold is set at 3. The spaxels where the S/N of at least one emission line is below the corresponding threshold are masked. Moreover, for the redshifted component, the masks described in Sect. \ref{sect:evolution_fit} are applied. The last panel displays the LINER, AGN, composite, and star-forming excitation regions as defined by \citet{2001ApJ...556..121K}, \citet{2003MNRAS.346.1055K}, and \citet{2007MNRAS.382.1415S}. The grey contours refer to galaxies representative of the RCSED sample.}
    \label{fig:first_bpt}
\end{figure*}
\subsubsection{$\mathrm{H \alpha}$-based SFR}
We derive an extinction-corrected star formation rate (SFR) for each component based on the measurements from the double-component decomposition. We corrected the H$\alpha$ luminosity following \citet{2001PASP..113.1449C}: $L_{\rm int}(\rm H\alpha)$=$L_{\rm obs}(\rm H\alpha)10^{0.4k(H\alpha)E(B-V)}$, where $L_{\rm int}$ is the intrinsic luminosity, $L_{\rm obs}$ is the observed luminosity, $k(H\alpha)$ is the reddening curve at the H$\alpha$ wavelength, and E(B-V) is the dust extinction. To compute this last value, we used the H$\alpha$/H$\beta$ ratio and assumed an intrinsic value of 2.86 (corresponding to a temperature of $T=10^4\,\rm K$ and an electron density $n_e=10^2\,\rm cm^{-3}$ for a case B recombination; \citealt{2006agna.book.....O}). Our derived values based on the MaNGA data, SFR$_{\rm H\alpha,1} = 7.9 \, M_{\odot} \, \mathrm{yr^{-1}}$ and SFR$_{\rm H\alpha,2} = 2.3 \, M_{\odot} \, \mathrm{yr^{-1}}$, are consistent with the different estimates of the SFR, namely from \citet{2004MNRAS.351.1151B} and from a spectral energy distribution (SED) fitting based on UV and optical spectra performed by \citet{2016ApJS..227....2S}, which agree towards a total value above $10 \, M_{\odot} \, \mathrm{yr^{-1}}$.

\subsubsection{WISE-based SFR}
The SFR derived from Wide-field Infrared Survey Explorer (WISE) data provides $10.57 M_{\odot} \rm yr^{-1}$ \citep{2016ApJS..227....2S}. It is in excellent agreement with the previous H$\alpha$-based estimate, supporting the assumption that there is no significant underestimation of the star formation activity. As discussed later (Fig. \ref{fig:sfr_vs_stellar_mass}), the two galaxies lie on the star formation main sequence.

\subsubsection{Radio-based SFR}
Following \citet{2019ApJ...872..148C} and assuming the radio flux is due to star formation only, we also derived a radio-based SFR that could be as large as $ 160\,M_{\odot} \, \mathrm{yr^{-1}}$. As mentioned in Sect. \ref{sect:vla}, the slope computed from the four flux densities (FIRST, both bands of VLA, and VLASS) is very steep. Although this value of -1.3 is not sufficient to assess whether the emission originates from AGNs or star formation, it is typical of a synchrotron emission rather than a free-free one.
The discrepancy  between the optical-based and the radio-based SFR estimates is due to either a severe underestimation of the extinction in the optical estimate or a hidden AGN or Low-Ionisation Nuclear Emission-line Region (LINER) contributing to the radio flux. However, the good agreement between the optical H$\alpha$-based SFR and the infrared WISE-based SFR discussed above rules out the first hypothesis.

As displayed in the bottom-left panel of Fig. \ref{fig:gas_2comp_properties}, the peak at 1.5\,GHz (as well as the peak at 2.5\,GHz) is coincident with the second component $\mathrm{H \alpha}$ flux. Given the large SFR observed for the main galaxy, the radio flux due to star formation would be expected to peak on this galaxy. The fact that the radio flux is centred on the second component is a strong argument in favour of a hidden AGN and supports the notion that most of this radio flux (more than 90\%) is not due to star formation.

\subsubsection{Off-centred star formation}
For the main galaxy, the peak of star formation detected in $\mathrm{H \alpha}$ is off-centred with respect to the kinematic centre. The velocity dispersion peaks at the centre of the rotation pattern, but some large gas velocity dispersions are detected in the north-west part of the disc. Indeed, while the stellar velocities are in good agreement with gas velocities, the stellar velocity dispersions are shifted with respect to those observed for the gas.
Moreover, we cannot exclude that the extinction correction of $\mathrm{H \alpha}$ emission is not sufficient to trace the actual ionised gas emission in the centre of the main galaxy. We might thus miss star formation at its centre. The stellar velocity dispersion might also be affected by an important amount of dust in the centre, which would explain why we detect high values in the north-west region of the galaxy where there is less dust.
The presence of clumps and filament-like structures, observed in the dust extinction maps (Fig. \ref{fig:gas_2comp_properties}) and in the velocity dispersion map of the primary galaxy (Fig. \ref{fig:gas_stars_kinematics}), also suggests the presence of underlying velocity gradients due to recent gas accretion triggered by the interaction.

\subsubsection{Two main-sequence galaxies}
\label{sect:sfms}
\begin{figure}[ht]
    \centering
    \includegraphics[width=.48\textwidth]{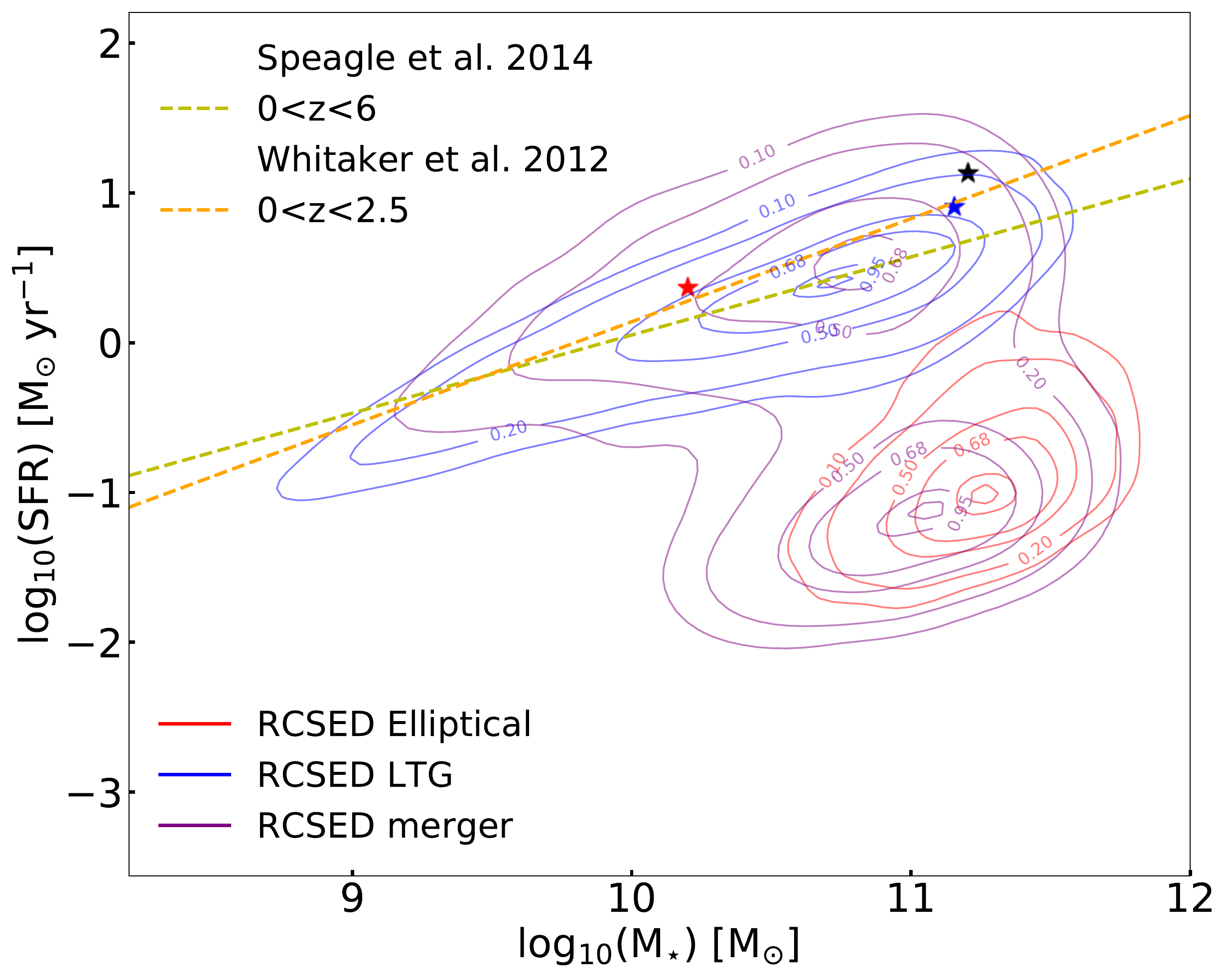}
    \caption{SFR vs. stellar mass plot. The black star corresponds to the total system, relying on the values provided by \citet{2016ApJS..227....2S} for J221024.49+114247.0. The blue and red stars refer to the main and secondary component, respectively, based on our SFR and stellar mass computations. The contours correspond to representative galaxies of the RCSED sample \citep{2017ApJS..228...14C}, classified using \citet{2018MNRAS.476.3661D}.}
    \label{fig:sfr_vs_stellar_mass}
\end{figure}
Figure \ref{fig:sfr_vs_stellar_mass} summarises the main star-forming properties of the two galaxies. They have a regular star-forming activity as both lie on the main sequence. For comparison, we display the contours based on galaxies taken from the RCSED \citep{2017ApJS..228...14C}. This catalogue encompasses 800,299 low- and intermediate-redshift galaxies ($0.007<z< 0.6$) and provides multi-wavelength spectrophotometric measurements from various surveys (GALEX, SDSS, and UKIDSS) as well as value-added data (k corrections, star and gas kinematics, and extinction estimates) for each object.
Globally, the two detected components currently do not exhibit any excess of star formation. This result is compatible with the molecular gas detection discussed in Sect. \ref{sect:CO}. 
Lastly, the excitation diagram explored next also supports the star-forming and composite excitations for most regions of these two galaxies.

\subsection{BPT excitation diagram}
When considering the Baldwin, Phillips \& Telervich diagnostic diagram \citep[introduced in][]{1981PASP...93....5B} for each peak separately (cf  Fig. \ref{fig:first_bpt}), most of the spaxels lie in the composite and star-forming regions for both peaks, but the second peak displays a systematically higher [OIII]/$\mathrm{H \beta}$ ratio than the first one. This supports our two-component analysis and is in line with the fact that the redshifted component has a smaller mass \citep[and lower metallicity; e.g.][]{2019ARA&A..57..511K,2020MNRAS.491..944C}. While  the two discs are mostly star forming, the BPT maps (Fig. \ref{fig:first_bpt}) display very localised LINER-type emission.
\begin{figure}[h]
    \centering
    \includegraphics[width=0.48\textwidth]{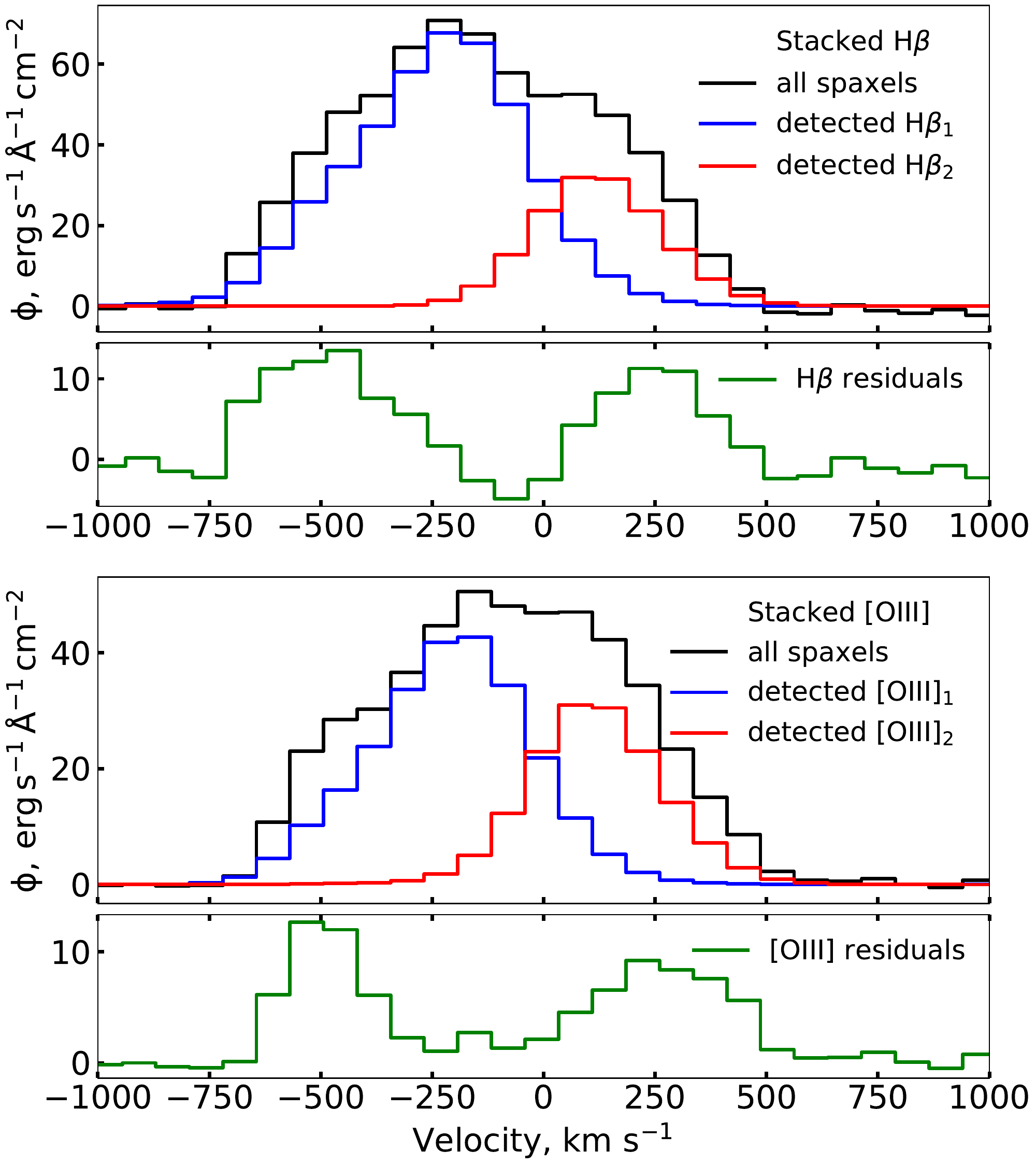}
   \caption{Stacked ionised-gas lines. In the top and bottom panels, we show the stacked H$\beta$ and [OIII]$\lambda$5008 emission lines, respectively, from all MaNGA spaxels in black. We show with blue and red the stacked blueshifted and redshifted components of our fitted model, only including spaxels with detected lines. Below the stacked emission lines, we show the residuals that exhibit excesses at large velocities.} 
    \label{fig:manga_stacked_lines}
\end{figure}

Based on the [OIII]$\lambda$5008 emission, it appears that the double-peaked profile for this line is reliably detected in a wider part of the MaNGA field of view. The spaxels identified as the secondary component do not seem to be associated with the main velocity gradient (their velocity is below $0\,\textrm{km}\ \textrm{s}^{-1}$) and show a higher excitation compared to the other spaxels related to this galaxy. This is reflected in their position on the BPT diagram, slightly shifted with respect to the main distribution of the other points and clearly belonging to the composite region (or even to the LINER region for three of them).
The integrated [OIII]$\lambda$5008 luminosities for each component within an area equivalent to the PSF (a circular region of about 2.5$^{\prime\prime}$) around the respective emission peak are $L_{[OIII],1}=1.1\,10^7L_{\odot}$ and $L_{[OIII],2}=8.3\,10^6L_{\odot}$. Even though we applied an extinction correction to derive these values, this might not be sufficient to trace the total [OIII]$\lambda$5008 emission in these inner parts of the galaxies. The values we find are higher than $10^{6.5}\,L_{\odot}$, the limit between weak and powerful AGNs considered by \citet{2007MNRAS.375.1017A}, which is consistent with their conclusion that most of the AGNs in pairs show an [OIII] luminosity above this threshold.

In Fig. \ref{fig:manga_stacked_lines} we also study stacked histograms of [OIII] and H$\beta$ fluxes. We both stacked the spectra of all spaxels and stacked the models of the blue and red components. Interestingly, we find some high velocity residuals that escaped the adjustment. They typically correspond to expected velocity fields in the outskirts of both discs, where the S/N is weak.


\subsection{Possible evidence of a radio jet}
Above we stated that the radio flux seems to be associated with the smaller galaxy. The radio emission from VLA observations is extended: We computed an integrated flux, called $F_{\rm int}$ hereafter, by summing up the emission over an area of 30x30 pixels around the flux peak and divided the derived value by the beam surface. We finally find $F_{\rm int}/F_{\rm peak}=8.7$ (respectively 5.1) at 2.5\,GHz (respectively 1.5\,GHz) with a 0.65$^{\prime\prime}$ (respectively 1.3$^{\prime\prime}$) resolution. Given the radio flux excess with respect to the expected SFR, we could argue that this is a weak radio jet.

\subsection{Molecular gas content}
\label{sect:CO}
The single-dish CO observations reveal an important reservoir of molecular gas within the system, both in CO(1-0) and CO(2-1). The inferred total H$_2$ mass represents 27\% of the stellar mass of the system. The Kennicutt-Schmidt relation shows that the system is located on the main sequence.

When we discriminate the flux below 0 $\rm km s^{-1}$ from the one above this value, as for the optical data, we find that the galaxy detected in the negative velocities has a higher CO(1-0)/CO(2-1) ratio than the secondary component. The beam sizes are different for both transitions, but since the CO(2-1) covers $\sim23\,\rm kpc$, we consider that we do not miss any significant emission outside the beam coverage. Thus, we can associate this ratio discrepancy with different excitations.

\subsection{Gas metallicity and post-starburst signature}
The maps of the gas metallicity are shown in the right panels of Fig. \ref{fig:gas_2comp_properties}. The smaller component appears to have a lower gas-phase oxygen abundance than the main galaxy, but this is not so clear when one computes averaged metallicities 12+log(O/H)$_1$=8.65 and 12+log(O/H)$_2$=8.58. However, the spatial distribution is very different. While the small component has a metallicity gradient roughly centred on the kinematics, the main galaxy exhibits a large gas metallicity region east of the star formation peak. This disordered metallicity gradient can be related to the gas accretion from the low-Z companion. This galaxy follows the mass-metallicity relation when we consider this off-centred area, while it exhibits a lower-than-expected metallicity in the central parts, as explained in Sect. \ref{sect:mz_relation} and shown in Fig. \ref{fig:mass_met}. Moreover, the galaxy spectra in these off-centred regions exhibit a significant Balmer absorption line that is attributed to a stellar population of $\sim$1\,Gyr \citep[e.g.][]{2018MNRAS.477.1708P}, typical of post-starburst regions. On the contrary, the spectra extracted from the regions where the computed gas-phase oxygen abundance is below the solar metallicity do not show such an important Balmer absorption line. Typical spectra extracted from these two different types of regions are displayed in Fig. \ref{fig:spectra_with_extreme_met}. 
Unlike in \citet{2007MNRAS.377.1222G}, the galaxies are still forming stars. Also, the starburst is quite old (1\,Gyr) compared to the systems observed in, for example, \citet{2000ApJ...529..157P}.

However, we do not see explicit or strong signs of a recently formed stellar population in the optical spectrum, but the S/N of the data makes it difficult to provide definitive conclusions based on individual bins. Ultraviolet data with a good spatial resolution would help us conduct a qualitative analysis of these regions. 

The irregular metallicity distribution can be compared with other merging systems, such as NGC4038/4039.\ This system was studied by \citet{2020MNRAS.497.3860G}, who detected star-forming regions with gas more enriched than the rest of the galaxy.


\subsection{Mass ratio estimates}
\subsubsection{Stellar mass ratio}
\label{sect:mstar}
To determine the light fraction between both detected stellar components, we performed a modelling of the stellar LOSVD.
The recovered LOSVD (see Sect. \ref{sect:nonparam_losvd}) is approximated in each spatial bin by two Gaussian components. Each of these components is represented by a thin inclined exponential disc. We used arbitrary initial guesses for the different characteristics of these profiles and eventually found an optimal set of parameters to reproduce the modelled LOSVD.
We obtain a light fraction of 6.1 between the main component and the secondary one. Using stellar ages and metallicities derived with \textsc{NBursts}, as mentioned in Sect. \ref{sect:nonparam_losvd}, we determined the mass-to-light ($M/L$) ratios for the $r$-band filter: $M/L_{r,1}=0.63$ and $M/L_{r,2}=0.43$. Given that the light fraction is 6.1, the stellar mass fraction is as follows: $6.1 \, ( M/L_{r,1}) / ( M/L_{r,2}) = 9$. This mass ratio enabled us to infer stellar masses for each of the detected component: According to \citet{2016ApJS..227....2S}, the total stellar mass of the system is $1.59\times10^{11}  M_{\odot}$, which, when considering the stellar mass ratio described above, gives: $ M_{\star,1}=1.43\times10^{11}\, M_{\odot}$ and $ M_{\star,2}=1.59\times10^{10}\,M_{\odot}$.

\subsubsection{Dynamical mass ratio}
\label{sect:mdyn}
For each galaxy, the inclination angle computed in Sect. \ref{ssect:kinegeo} can be used to calculate the respective rotation velocity as in \citet{2014ApJ...795L..37C} and \citet{2018MNRAS.479.2133A} following ${V_{\rm rot} = W / 2 \, \sin(i)}$, where $W$ is defined as the difference between the 90th and 10th percentile points of the velocity histogram.
We derive a rotation velocity of $ V_{\rm rot,1}=236\, \mathrm{km \, s^{-1}}$ for the main galaxy and $ V_{\rm rot,2}=140\, \mathrm{km \, s^{-1}}$ for the secondary one. Then we computed the dynamical mass corresponding to the first MaNGA component as follows: $ M_{\rm dyn,1}={V_{\rm rot,1}^2\,R_1}/{G}$, where $R_1$ is the maximal radius where $\mathrm{H \alpha}$ emission is detected. We find $ M_{\rm dyn,1}=2.65\times10^{11}\,M_{\odot}$.
Similarly, the dynamical mass of the second component is defined as: $ M_{\rm dyn,2}={V_{\rm rot,2}^2\,R_2}/{G}$. We derive $ M_{\rm dyn,2}=2.93\times10^{10}\,M_{\odot}$. These results give a mass ratio of $M_{\rm dyn,1}/M_{\rm dyn,2}\sim9$.

\subsubsection{Uncertainties}
To evaluate the robustness of the previous estimate of the merger mass ratio, we varied the parameters used in the computations of the stellar and the dynamical masses.
As the interaction between both objects certainly perturbs both galaxies, we can consider that it is relevant to include the contribution of the velocity dispersion $\sigma_{\rm gas}$ in the dynamical mass estimate. Based on the $S_{0.5}$ parameter defined and discussed in \citet{2018MNRAS.479.2133A}, we calculated the dynamical mass using the proxy they introduce as $ M_{\rm dyn}={\eta\,R\,S^2_{0.5}}/{G}$. The inferred dynamical masses are $ M_{\rm dyn,1}=6.98\times10^{11}\,M_{\odot}$ and $ M_{\rm dyn,2}=5.80\times10^{10}\,M_{\odot}$, corresponding to a mass ratio $M_{\rm dyn,1}/M_{\rm dyn,2}$ of 12.
In parallel, for the stellar masses, we can force both inclinations of the modelled exponential disc profiles based on the inclination angles computed using the gas velocity fields. This leads to a light fraction of 7.3. Relying on the mass-to-light ratios given in Sect. \ref{sect:mstar}, the corresponding stellar mass ratio is $\sim 10$.
Lastly, one can note that considering a large inclination angle for the secondary galaxy provides a dynamical mass similar to the stellar mass, constraining an inclination smaller than 80$^\circ$.


Finally,
all the approaches converge on a mass ratio within the visible radius of about 9:1 for MaNGA 1-114955.

\subsubsection{Mass-metallicity relation}
\label{sect:mz_relation}
\begin{figure}[ht]
    \centering
    \includegraphics[width=.48\textwidth]{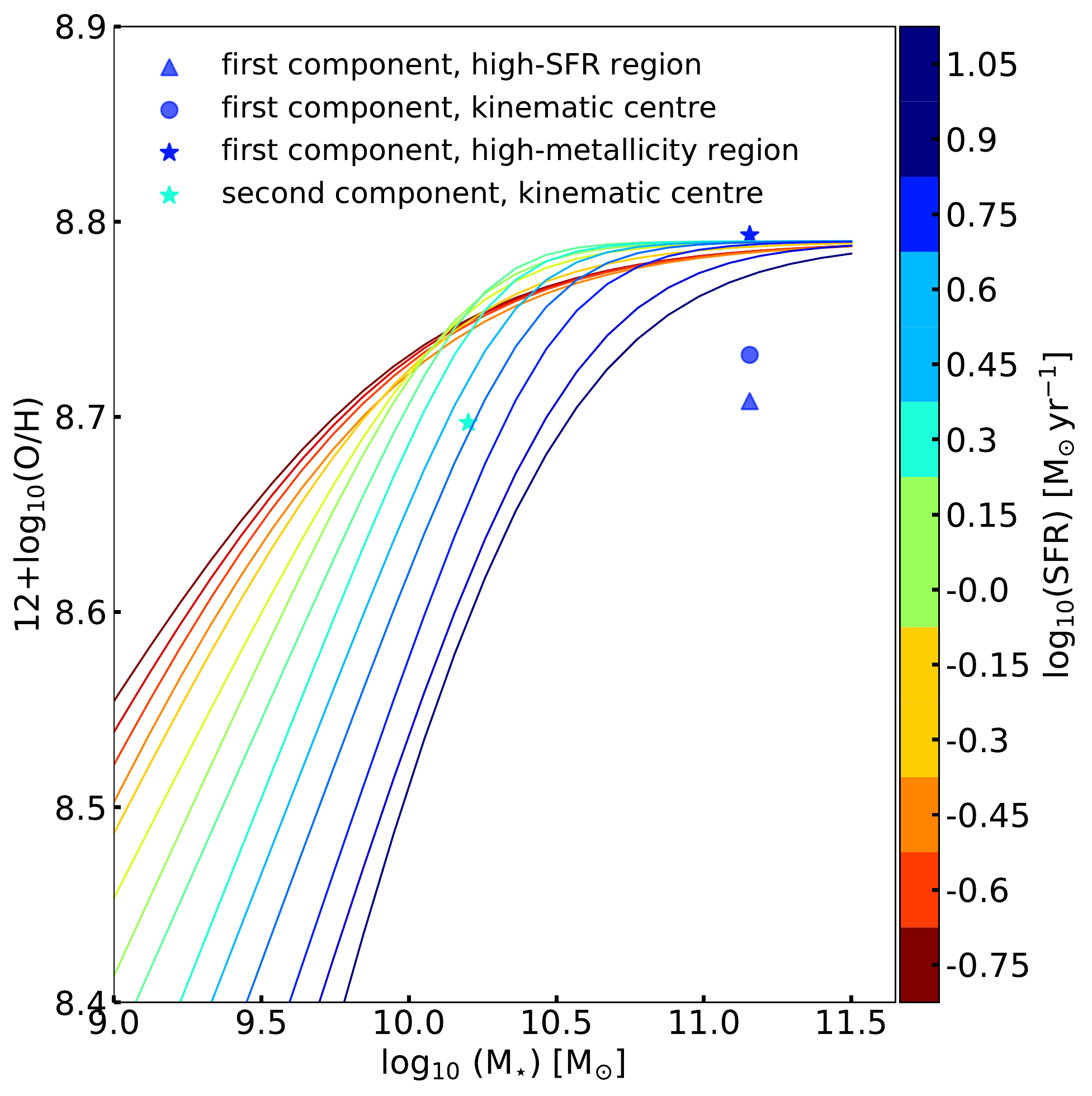}
    \caption{Mass-metallicity relations for different values of total SFR, as defined in \citet{2020MNRAS.491..944C}. The SFRs of both components are taken from \citet{2016ApJS..227....2S}, and the stellar masses are the ones given in Sect. \ref{sect:mstar}.The markers are colour-coded with the SFR of the corresponding host galaxies. For the primary galaxy, different 3 arcsec circular apertures have been considered, as discussed in Sect. \ref{sect:mz_relation}.}
    \label{fig:mass_met}
\end{figure}

In order to determine if the metallicities of these two galaxies, which both lie on the star formation main sequence, are perturbed, we tried to position them on the stellar mass-SFR-gas metallicity relation proposed by \citet{2020MNRAS.491..944C} \citep[based on][]{2010MNRAS.408.2115M}.
Figure \ref{fig:mass_met} shows the mass-metallicity relations for different SFR bins as described in \citet{2020MNRAS.491..944C}. It should be noted that the gas metallicity computed here corresponds to the gas metallicity computed in the SDSS 3 arcsec (central) fibre. We used the calibration defined in this work to compute the metallicity, based on the $O_3N_2$ calibrator, within circular regions with a diameter of 3 arcsec. On the one hand, the second galaxy lies almost on the curve parametrised for objects with a similar SFR. On the other hand, the properties of the first galaxy are consistent with the relation only if we compute the metallicity over the off-centred region with high gas-phase oxygen abundance values (see the top-right panel of Fig. \ref{fig:gas_2comp_properties}). If we compute the metallicity around the kinematic centre or the star formation peak, the value is too low compared to the mass of the galaxy and thus does not follow the relation. We can argue that this primary galaxy has probably accreted metal-poor gas in its centre through the interaction.

\section{Discussion}
\label{sect:disc}

\subsection{Pre-coalescence merger}
\begin{figure*}[ht]
    \centering
    \includegraphics[width=.89\textwidth]{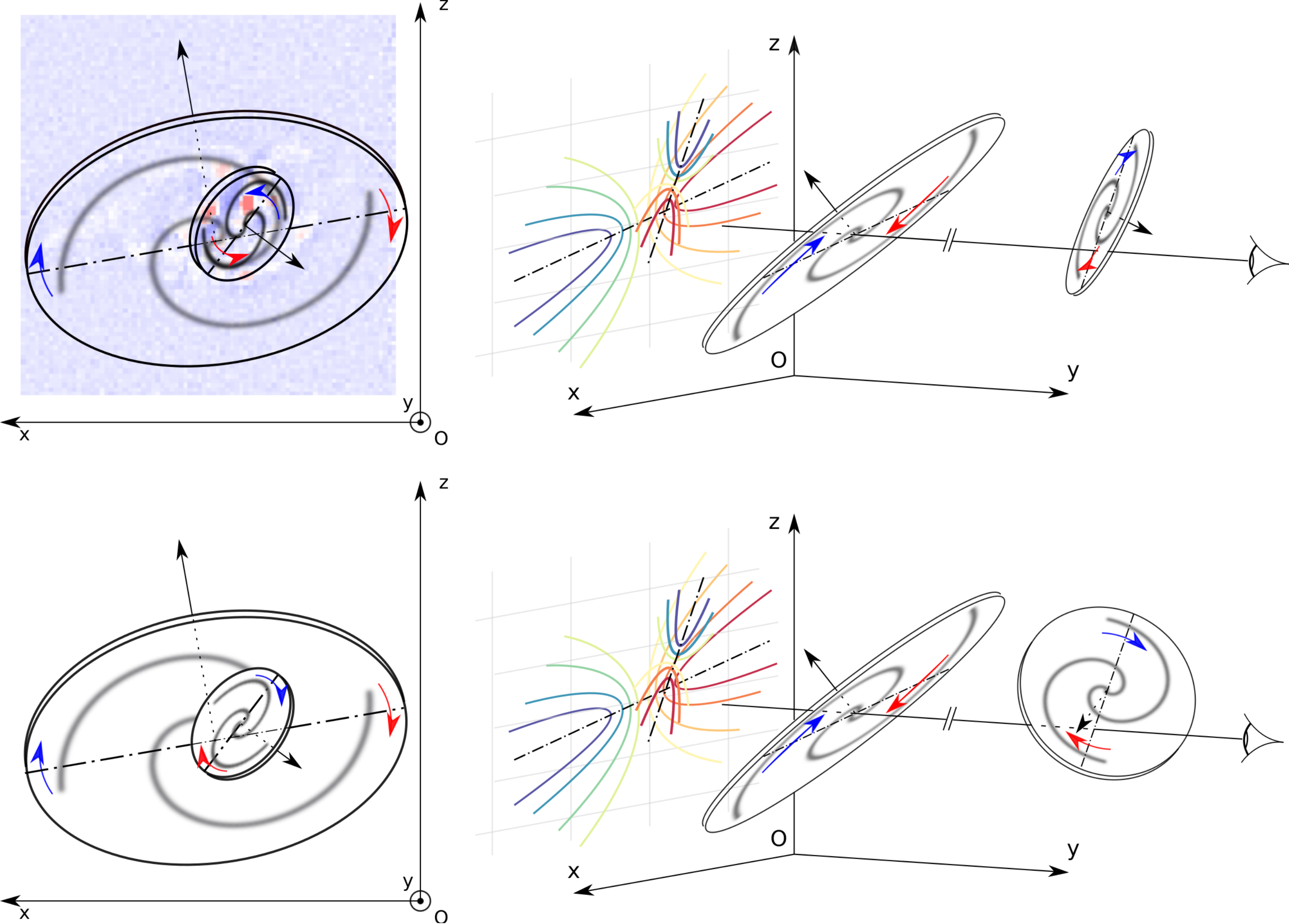}
    \caption{Schematic illustration of the revealed merging system. \textbf{Top:} Configuration where the near side of the secondary galaxy is on the north side.  \textbf{Bottom:} Configuration where the near side is on the south side of the major axis. The left panels show the discs as projected on the sky. The right panels show a `3D view', with a different viewing angle, for which discs can be represented with no superposition. The top-left panel includes the highlighting of a potential dust lane in Fig.~\ref{fig:rgb_legacy_rdiff_megapipe}.}
    \label{fig:schematic_view}
\end{figure*}

We identify J221024.49+114247.0 as a merging system caught in action. We observe two distinct counter-rotating  discs, aligned along the line of sight, with a line-of-sight velocity difference of $ \Delta V = 450 \, \mathrm{km\,s^{-1}}$. We find one main galaxy with an inclination of $i_1 = 49\,^{\circ}$ and a second galaxy of $i_2 = 46\,^{\circ}$ as a smaller counterpart. 
Using high-resolution imaging from the Megacam-CFHT footprint, we identified a dust lane that is aligned with the stellar component of the smaller galaxy and is associated with measured extinction in the main galaxy, suggesting a scenario where the small galaxy is positioned in front of the main galaxy along the line of sight, as shown in Fig.\,\ref{fig:schematic_view}. 
These two disc components are both detected in the ionised gas and in the stellar continuum emission, and they have not yet been destroyed by the interaction. 
We estimated a mass ratio of 9:1 using the stellar masses and dynamical masses.
For both galaxies, we find off-centred peaks in stellar velocity dispersion, reminiscent of possible accretion onto the disc and an underlying filamentary structure.

%



We show schematic representations of two possible configurations of the observed objects in Fig.~\ref{fig:schematic_view}: Each row of the figure shows one of these configurations, viewed as projected on the sky (left panels) or with a viewing angle (the same for the two rows) for which the discs can be represented with no superposition (right panels, '3D views'). The line of sight along which the system is observed with MaNGA is parallel to the (Oy) axis (the MaNGA observer is symbolised with the eye to the right). The disc on the left of the 3D views represents the primary galaxy, and the one on the right, the secondary. The rotation of each disc is shown with a spin (black arrow) and corresponding schematic trailing spiral arms. The coloured arrows represent the line-of-sight velocity for the MaNGA observer (blue- or red-shifted with respect to each kinematic centre), and the MaNGA velocity maps are schematically represented on the (Oxz) plane of the right panels (with different colours corresponding to each kinematic centre, as in Fig.~\ref{fig:gas_stars_kinematics}). We note that the distance between the two galaxies is unknown (hence the break in the line of sight). For a given disc inclination, part of one side of the major axis of the projection of the disc on the sky is either farther away than this major axis or closer to it (near side). We can determine the near side of the primary galaxy with the help of its spiral arms, seen in Fig.~\ref{fig:rgb_legacy_rdiff_megapipe}. We assumed that these spiral arms are trailing, which gives the sense of rotation on the sky. Combined with the line-of-sight velocities, this determines the inclination of the disc plane, with a near side north-east of the major axis. The near side of the secondary galaxy is harder to determine. The presence of the dust lane parallel to the kinematic axis but shifted to the north could support an argument that the near side is on the north side, as in the top panel of Fig.~\ref{fig:schematic_view}. Alternatively, the bottom panel shows a configuration compatible with the observation of an ionisation cone north-east of the major axis of the secondary disc, which would mean that the near side is on the south side of the major axis.

\subsection{Possible outflows driven by central activity}
The interaction might also have triggered gas accretion on the supermassive black hole (SMBH) and a hidden AGN detected in radio. This is compatible with the discussion in \citet{2020ApJ...904..107S} regarding the possibility of low-level AGN enhancement in galaxy interactions. Using BPT diagrams, we find that the two galaxies are located in different regions, associated with different states of ionisation. On the one hand, the star-forming and composite excitation occupies different regions, probably due to the different galaxy masses, while the second galaxy shows some regions with LINER excitation, suggesting a different mechanism that is potentially linked to the radio flux found to be associated with this galaxy. This might indicate that the merger process can induce different mechanisms, shifting the two components on the BPT diagram, as discussed in \citet{2019A&A...627L...3M}.

\begin{figure}[ht]
    \centering
    \includegraphics[width=.48\textwidth]{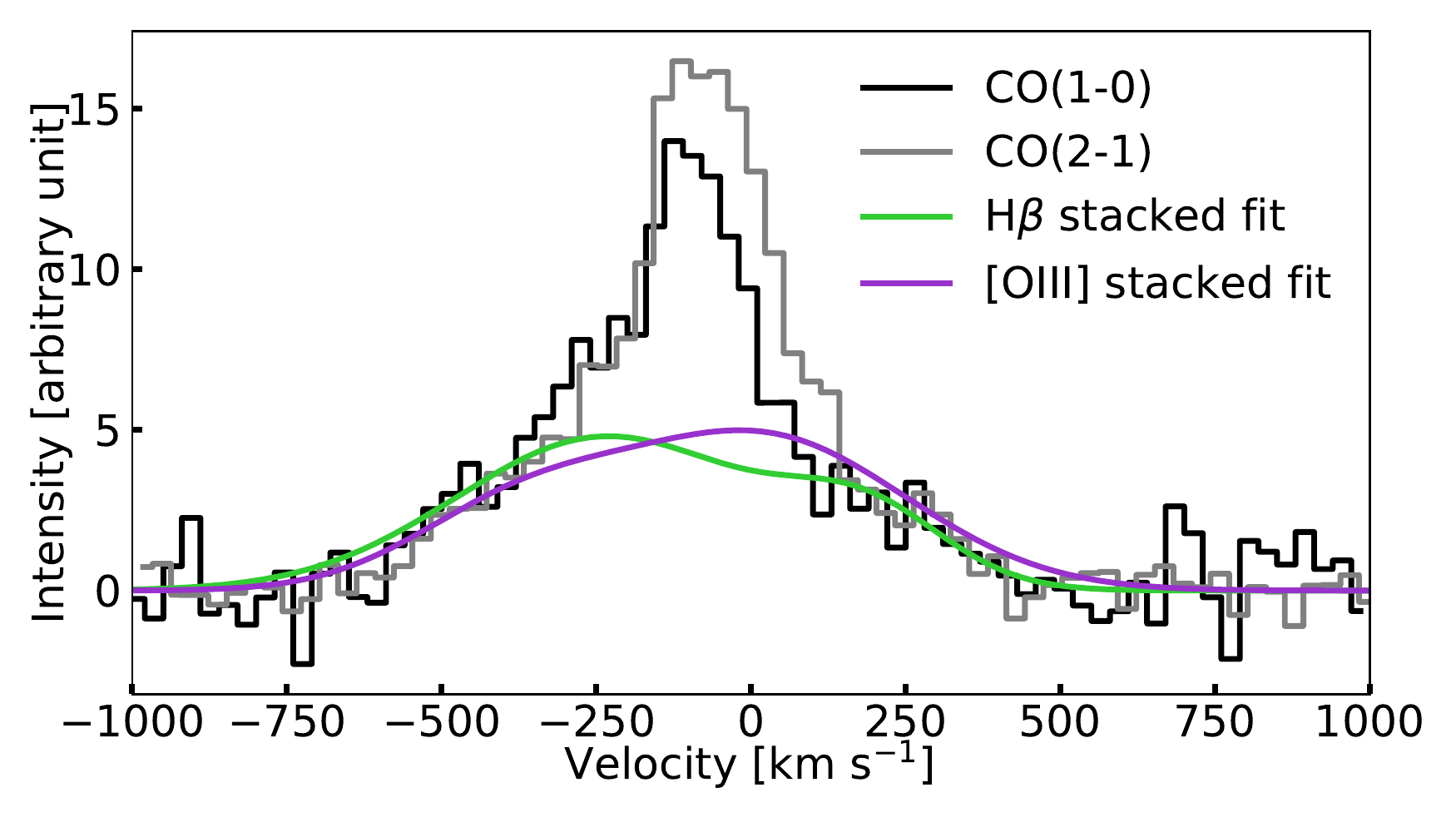}
    \caption{Re-scaled CO spectra and fits of the stacked H$\beta$ and [OIII] lines that display a similar range of velocities.}
    \label{fig:manga_gaussians_both_co}
\end{figure}

The molecular gas spectra obtained for both galaxies exhibit a large range of velocities of about $1000\,\rm km\, s^{-1}$. By comparing these spectra to the stacked H$\beta$ and [OIII] emission lines from the optical spectra in Fig. \ref{fig:manga_gaussians_both_co}, we observe that the wide wings of CO emission coincide with the wings observed in the ionised gas. This accounts for the rotation patterns of the two discs. In parallel, we detect, in the velocity range -100-0\,$\rm km\, s^{-1}$, a relative excess of molecular gas corresponding to the peak of star formation activities. To understand how the molecular gas behaves in the vicinity of the possible AGN identified in radio, higher-resolution observations of molecular gas with the NOrthern Extended Millimeter Array (NOEMA) are needed.

The wide range of velocities detected in both molecular and ionised gas might correspond to a gaseous outflow, either due to the star formation triggered by the interaction or to the feedback due to the AGN in the small galaxy. It has been shown that radio jets, even when compact, can create major turmoil in galaxies, enhancing the velocity width \citep{Venturi2021}.
Figure \ref{fig:manga_stacked_lines} reveals that the wide wings are not detected in individual spectra, but appear only by stacking. This turbulent or outflowing gas must be spread over some diffuse regions and only at low intensity. 

\subsection{Evidence of two pericentre passages}
The total SFR of the entire system is estimated to be $13.3\,{M_{\odot}\,\rm yr^{-1}}$ \citep{2016ApJS..227....2S}.
Radio observations suggest a much higher SFR of more than $80\,{M_{\odot}\,\rm yr^{-1}}$, which is probably biased by a contribution of an obscured AGN. 
Using the extinction-corrected H$\alpha$ emission line, we located a concentration of ongoing star formation in the main galaxy, which is off-centred with respect to the kinematic centre by about 5\,kpc in projection.
The smaller galaxy exhibits a strong peak in star formation, located at its kinematic centre, and is associated with an extended radio emission.

It seems that the external gas of the smaller companion has been stripped by the main galaxy, which has probably accreted this gas onto its disc. It is hence probable that the progenitor of the smaller galaxy was originally more massive. The ignition of the AGN, somehow related to the gas present in the central starburst, is due either to accretion of gas (from the primary) or to the original central distribution of gas, and is not necessary related to the harassment. 
We also observe a peak in the gas-phase metallicity of the main galaxy located to the east of its main star formation region. This high-metallicity region, located more than 6\,kpc away from the kinematic centre, is associated with significant Balmer absorption lines imprinted in the galaxy spectrum (see Fig. \ref{fig:spectra_with_extreme_met}). This feature could trace past starburst events not older than 1\,Gyr \citep[e.g.][]{2018MNRAS.477.1708P}. Indeed, this clearly asymmetric metallicity distribution is not expected in a relaxed system. We argue that this could be the signature of a first pericentre passage, while the current star formation (in the north-west part of the main galaxy) could be associated with a recent pericentre passage.
Star formation induced by the interaction of galaxies varies significantly (in terms of highest SFR, duration of high SFR periods, and integrated star formation on the duration of the merger) as a function of the orbital parameters and the inclination of the discs of the progenitors \citep[e.g.][]{2008A&A...492...31D}. It is thus difficult to infer the geometry of the interaction (prograde or retrograde orientation of the spins of the galaxies) from the observed star formation.

\subsection{Comparison with other merging systems}
The interacting galaxies presented in this work can be compared to the less massive system Mrk\,739, which hosts a double AGN \citep{2021ApJ...911..100T}. In this case, two AGNs have been identified in X-ray \citep{2011ApJ...735L..42K}. Interestingly, in this source the ionised gas distribution clearly peaks at the position of the two AGNs. The gas is concentrated in the centre, which explains the ignition of both AGNs. As the system is more face-on with a projected nuclear separation of 3.4\,kpc, it is not obvious if there are two rotating distinct discs, such as those we observe for J221024.49+114247.0. The authors noted that the progenitors were a star-forming galaxy and an old elliptical. 

In our case, the main galaxy has a larger SFR than its companion, but this is probably triggered by accretion of the stripped gas. Interestingly, this gas did not reach the galaxy centre and was not able to fuel the SMBH.  Using a simulation of NGC\,2623, \citet{2021arXiv210109407P} explained that the first passage can trigger extended star formation, such as we observed here. \citet{2020MNRAS.496.5243H} discussed the two galaxies in Arp\,240 with a projected distance of about 40\,kpc. They claim that the low depletion time (100\,Myr) is an argument for a late-stage merger. In the case of J221024.49+114247.0, the global depletion time (3.3\,Gyr) is typical of normal star-forming galaxies \citep[e.g.][]{2011MNRAS.415...61S}.


\section{Conclusions}
We have presented a multi-wavelength analysis of a pre-coalescence merger, J221024.49+114247.0, at $z=0.09$. Originally identified with a double-peaked feature in the SDSS fibre and a disrupted merger morphology, this galaxy has been observed with MaNGA. We developed a fitting procedure based on a double-Gaussian function and applied it throughout the MaNGA field of view. We have been able to highlight the presence of two counter-rotating discs along the line of sight in gas as well as in stars. The difference of their systemic velocity is 450\,km\,s$^{-1}$, and they are aligned along the line of sight. A modelling of the stellar LOSVD enables us to derive a stellar mass ratio between the main and the secondary component of $\sim 9$, relying on the mass-to-light ratio estimates. The mass ratio based on dynamical mass estimates is in good agreement with that derived from the stellar distribution analysis.

The main galaxy exhibits off-centred regions of star formation. A large region (at $\sim 6 \, {\rm kpc}$ from the kinematic centre) exhibits a large gas metallicity and a Balmer absorption line in the galaxy spectra, possibly reminiscent of a 1\,Gyr old starburst, while the current SFR detected in $\mathrm{H \alpha}$ is also off-centred with respect to the kinematic centre. We argue that this is due to two different pericentric passages. This more massive galaxy is probably more gas rich than its companion, but the gas has not collapsed in the centre. The minor galaxy exhibits an excess of extended radio flux centred on its kinematic centre with respect to the expected SFR. Its kinematic centre corresponds to a standard velocity dispersion and metallicity gradient. We argue that its external gas has been stripped. Large molecular gas velocities are compatible with the ionised gas rotation velocities, and we cannot conclude on the possibility of an outflow.

We propose extending this work with NOEMA/IRAM interferometric observations in order to study the molecular gas distribution and the sub-millimetre continuum distribution.




\begin{acknowledgements}
We are most grateful to the IRAM-30m team who supported us for these observations.
IK acknowledges the support from the Russian Scientific Foundation grant 19-12-00281 and the Interdisciplinary Scientific and Educational School of Moscow University ``Fundamental and Applied Space Research''.
We thank the anonymous referee for constructive comments.\\
Funding for the Sloan Digital Sky Survey IV has been provided by the Alfred P. Sloan Foundation, the U.S. Department of Energy Office of Science, and the Participating Institutions. SDSS-IV acknowledges
support and resources from the Center for High-Performance Computing at the University of Utah. The SDSS web site is www.sdss.org.
SDSS-IV is managed by the Astrophysical Research Consortium for the 
Participating Institutions of the SDSS Collaboration including the 
Brazilian Participation Group, the Carnegie Institution for Science, 
Carnegie Mellon University, the Chilean Participation Group, the French Participation Group, Harvard-Smithsonian Center for Astrophysics, 
Instituto de Astrof\'isica de Canarias, The Johns Hopkins University, Kavli Institute for the Physics and Mathematics of the Universe (IPMU) / 
University of Tokyo, the Korean Participation Group, Lawrence Berkeley National Laboratory, 
Leibniz Institut f\"ur Astrophysik Potsdam (AIP),  
Max-Planck-Institut f\"ur Astronomie (MPIA Heidelberg), 
Max-Planck-Institut f\"ur Astrophysik (MPA Garching), 
Max-Planck-Institut f\"ur Extraterrestrische Physik (MPE), 
National Astronomical Observatories of China, New Mexico State University, 
New York University, University of Notre Dame, 
Observat\'ario Nacional / MCTI, The Ohio State University, 
Pennsylvania State University, Shanghai Astronomical Observatory, 
United Kingdom Participation Group,
Universidad Nacional Aut\'onoma de M\'exico, University of Arizona, 
University of Colorado Boulder, University of Oxford, University of Portsmouth, 
University of Utah, University of Virginia, University of Washington, University of Wisconsin, Vanderbilt University, and Yale University.\\
The Legacy Surveys consist of three individual and complementary projects: the Dark Energy Camera Legacy Survey (DECaLS; Proposal ID \#2014B-0404; PIs: David Schlegel and Arjun Dey), the Beijing-Arizona Sky Survey (BASS; NOAO Prop. ID \#2015A-0801; PIs: Zhou Xu and Xiaohui Fan), and the Mayall z-band Legacy Survey (MzLS; Prop. ID \#2016A-0453; PI: Arjun Dey). DECaLS, BASS and MzLS together include data obtained, respectively, at the Blanco telescope, Cerro Tololo Inter-American Observatory, NSF’s NOIRLab; the Bok telescope, Steward Observatory, University of Arizona; and the Mayall telescope, Kitt Peak National Observatory, NOIRLab. The Legacy Surveys project is honored to be permitted to conduct astronomical research on Iolkam Du’ag (Kitt Peak), a mountain with particular significance to the Tohono O’odham Nation.\\
This research used the facilities of the Canadian Astronomy Data Centre operated by the National Research Council of Canada with the support of the Canadian Space Agency.\\

\end{acknowledgements}
\vspace{-1cm}
\bibliographystyle{aa}
\bibliography{biblio}

\clearpage
\newpage
\appendix
\section{Gas spatial properties obtained through our fitting procedure}
\label{sect:maps_double_gauss}
\subsection{Flux maps}
\begin{wrapfigure}{c}{0.98\textwidth}
   \includegraphics[width=.48\textwidth]{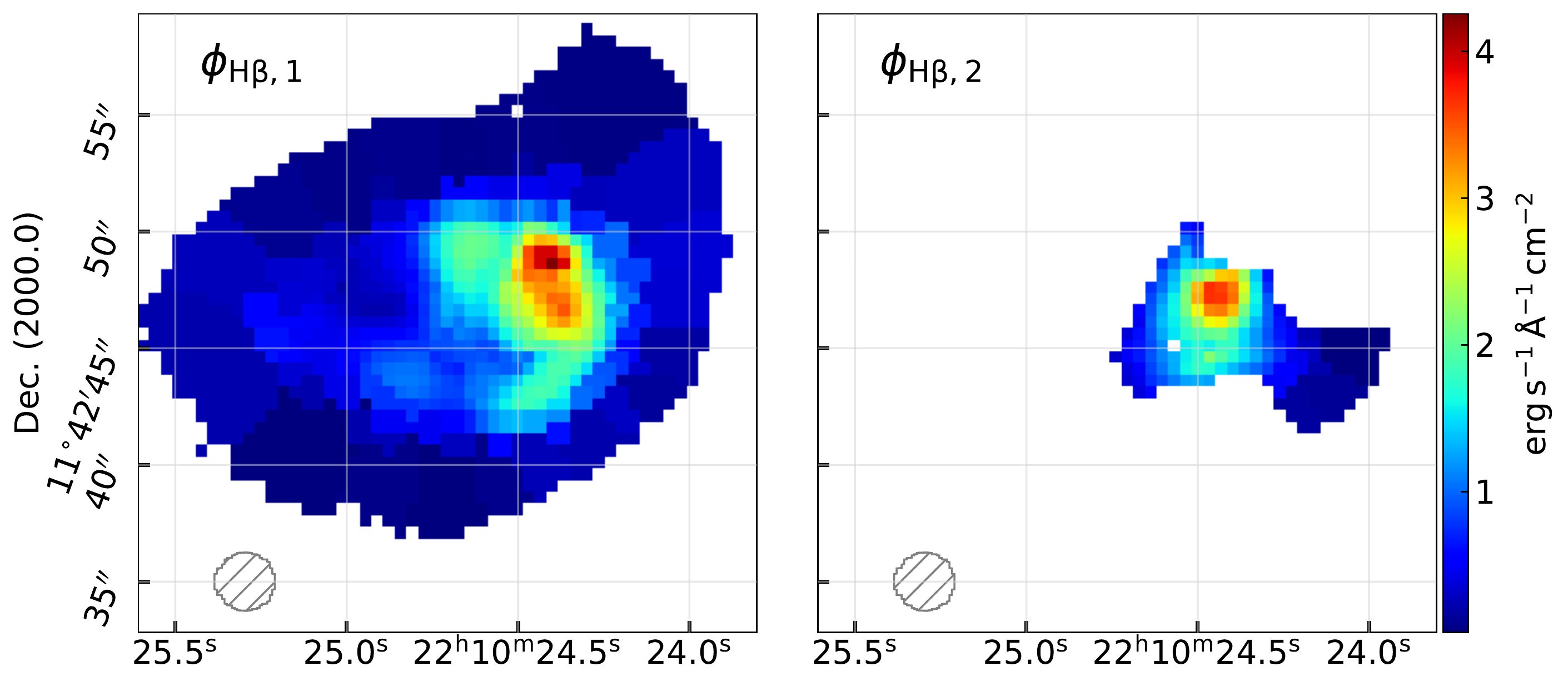}
   \includegraphics[width=.48\textwidth]{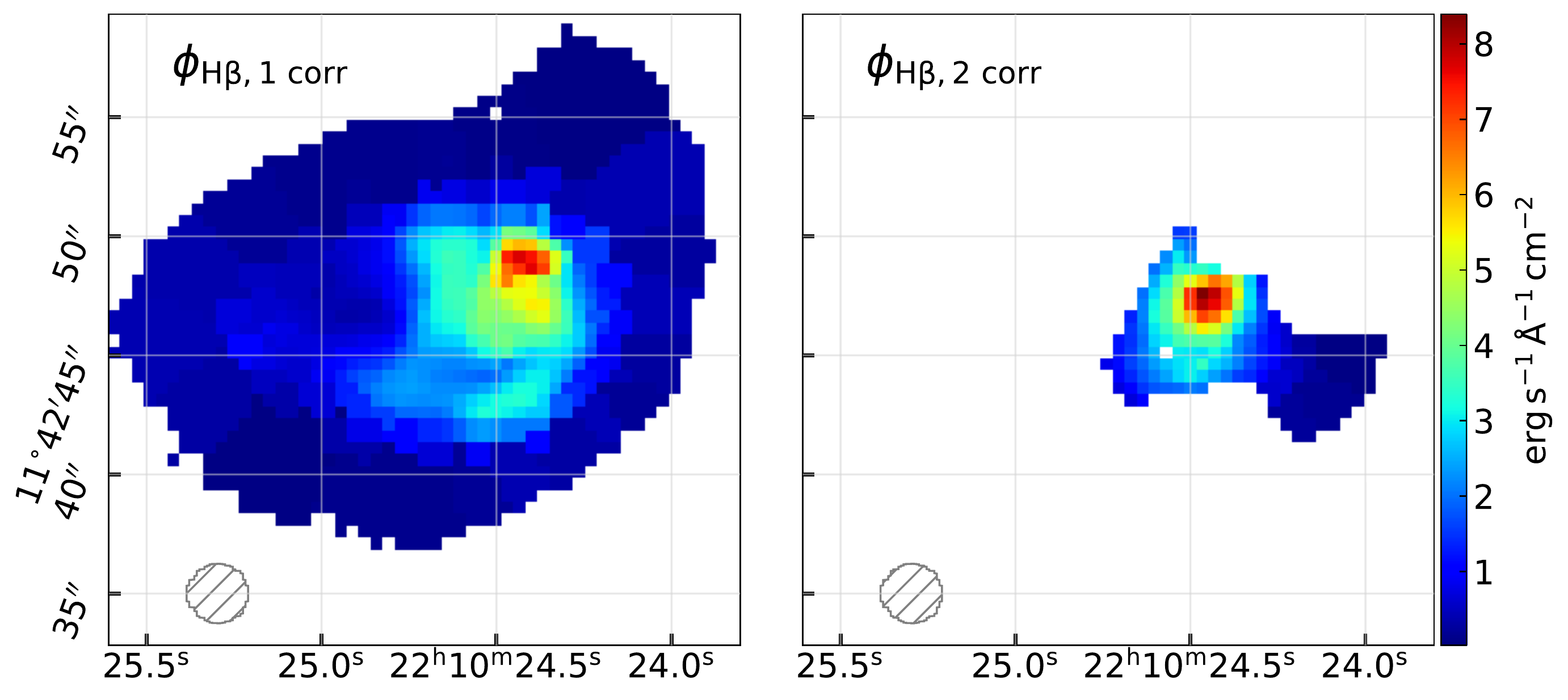}\\
   \includegraphics[width=.48\textwidth]{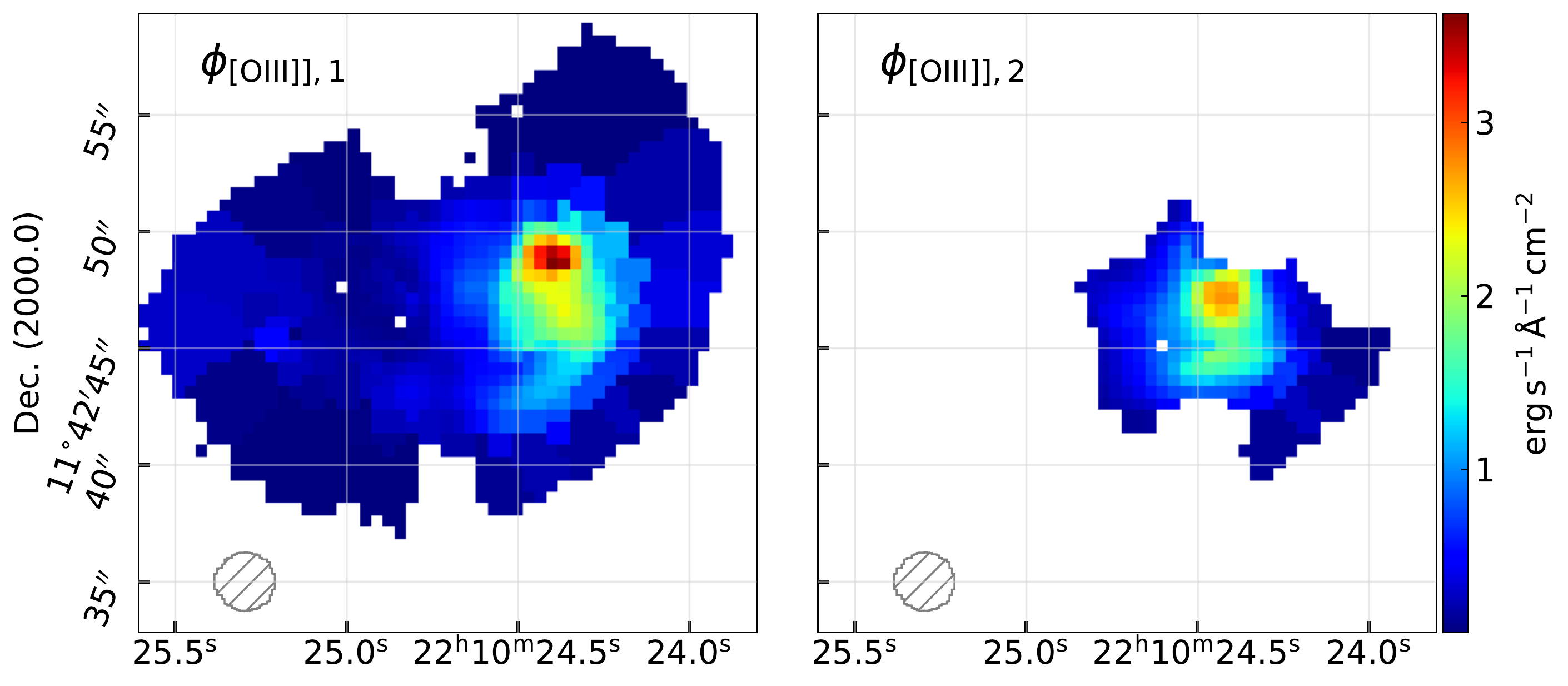}
   \includegraphics[width=.48\textwidth]{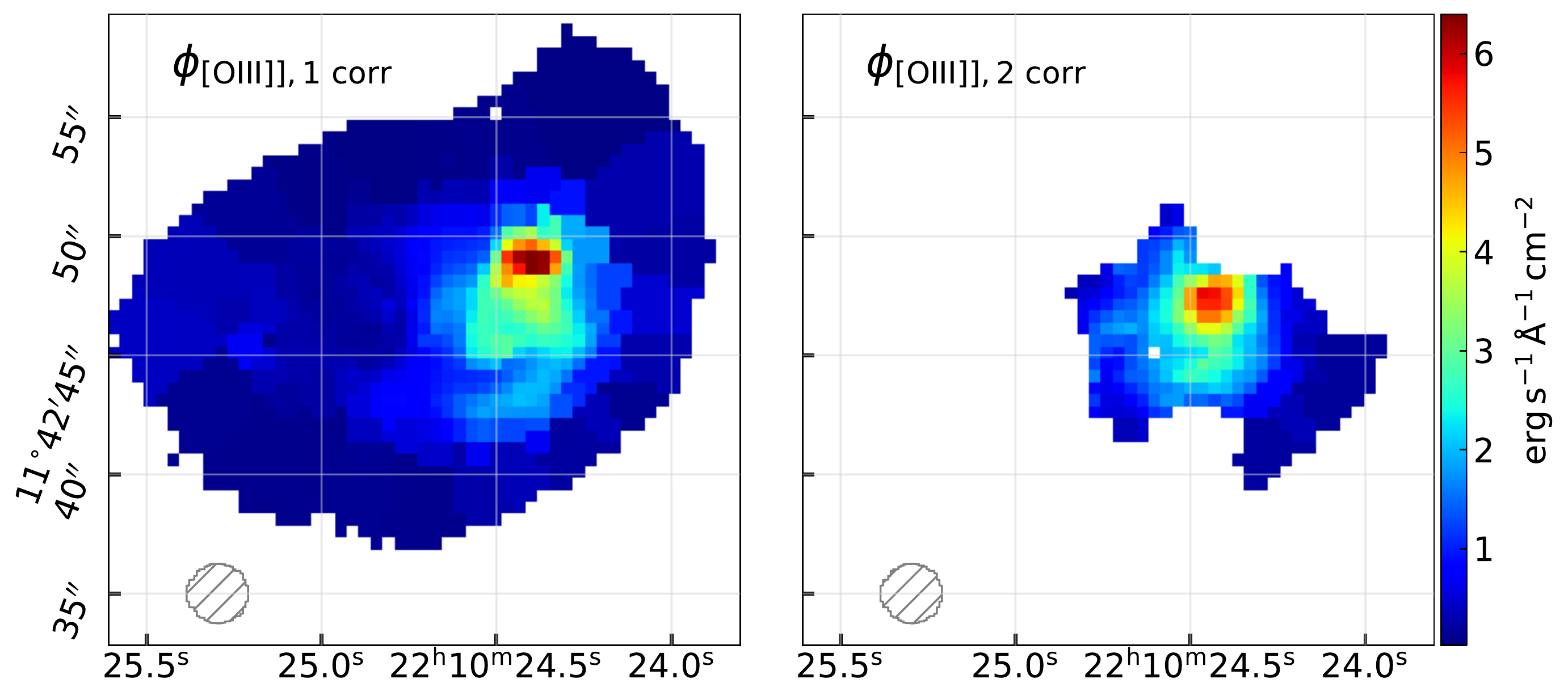}\\
   \includegraphics[width=.48\textwidth]{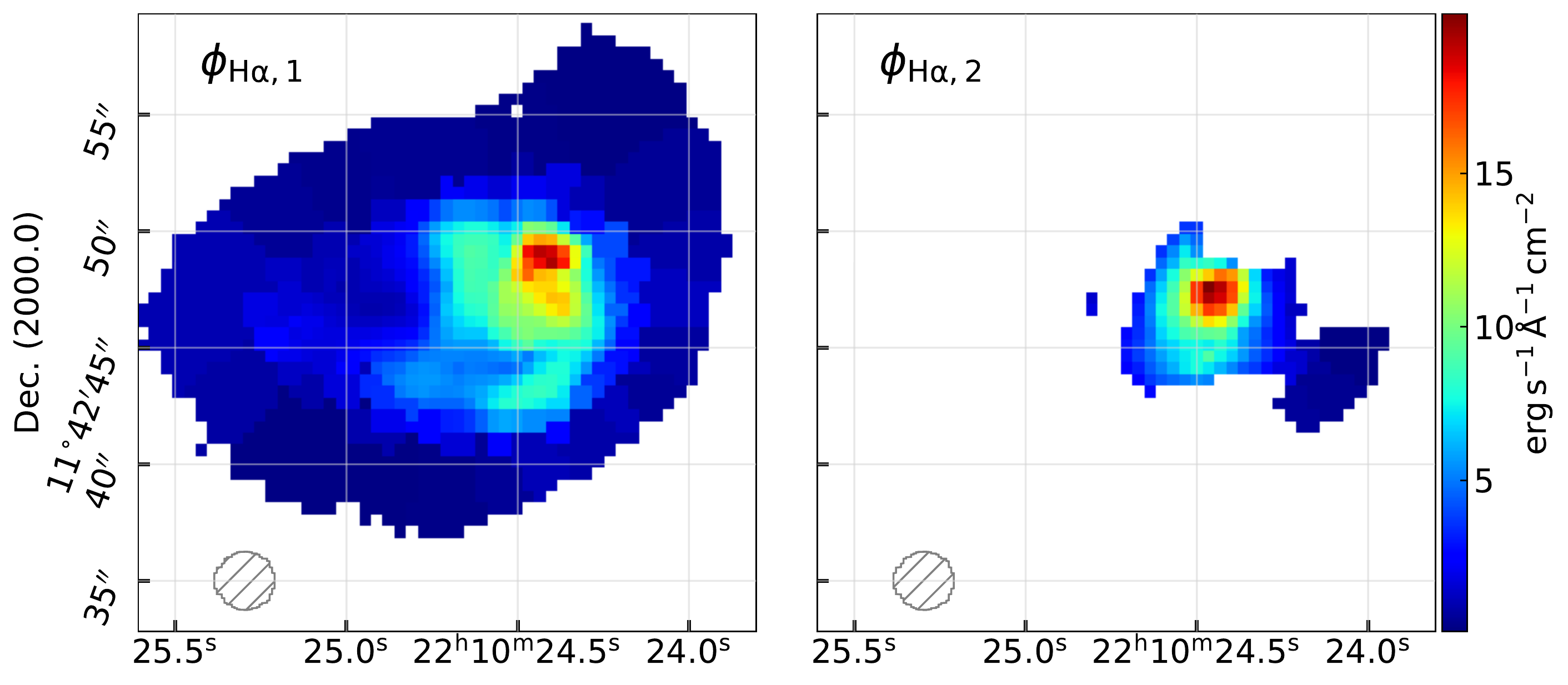}
   \includegraphics[width=.48\textwidth]{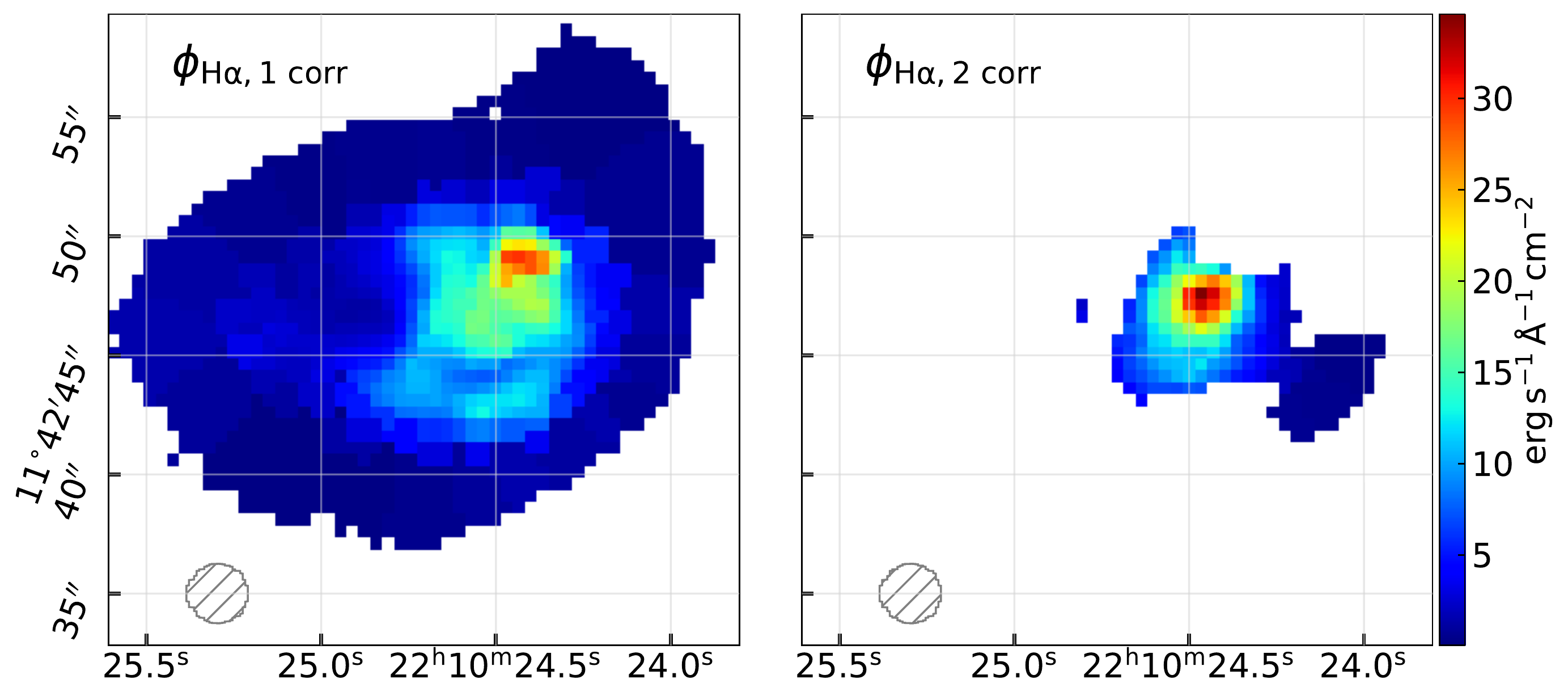}\\   
   \includegraphics[width=.48\textwidth]{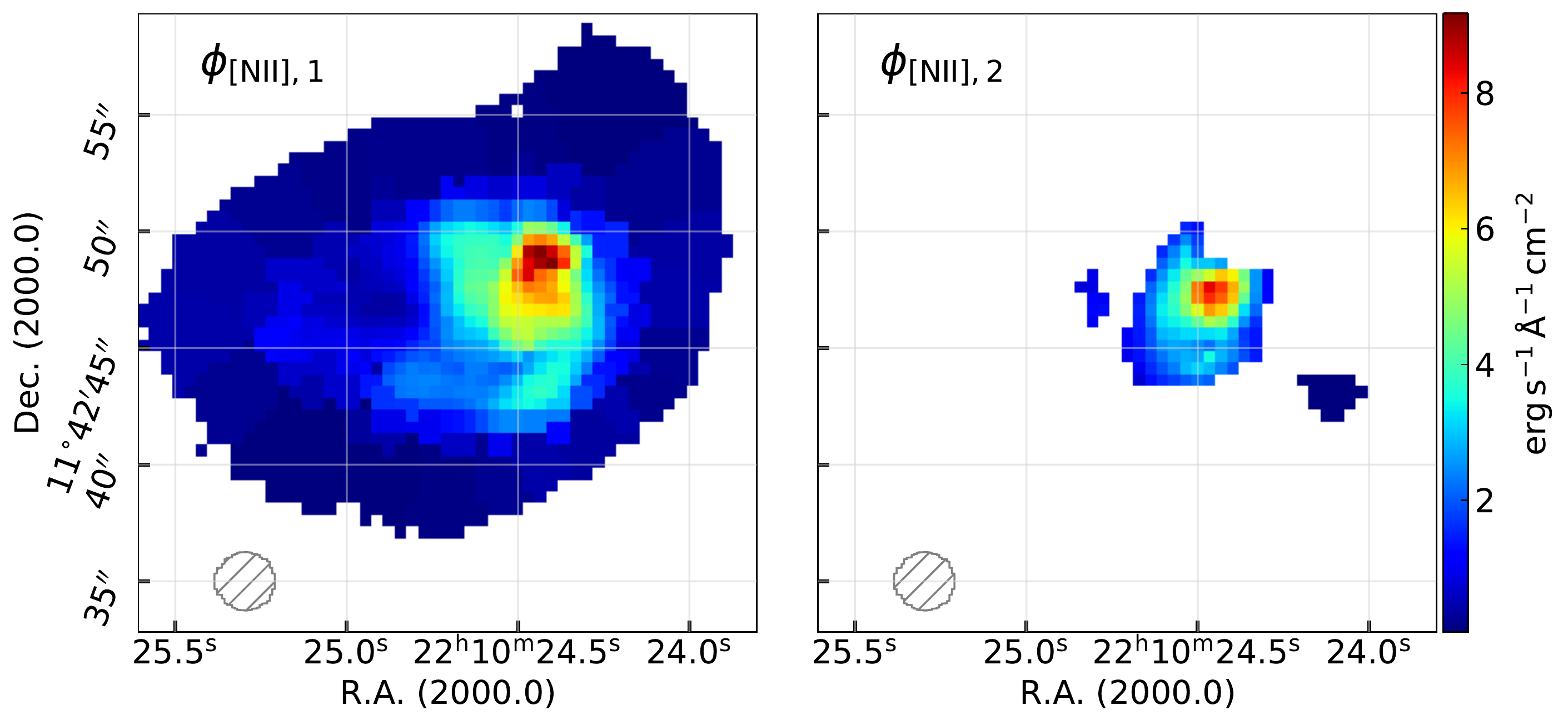}
   \includegraphics[width=.48\textwidth]{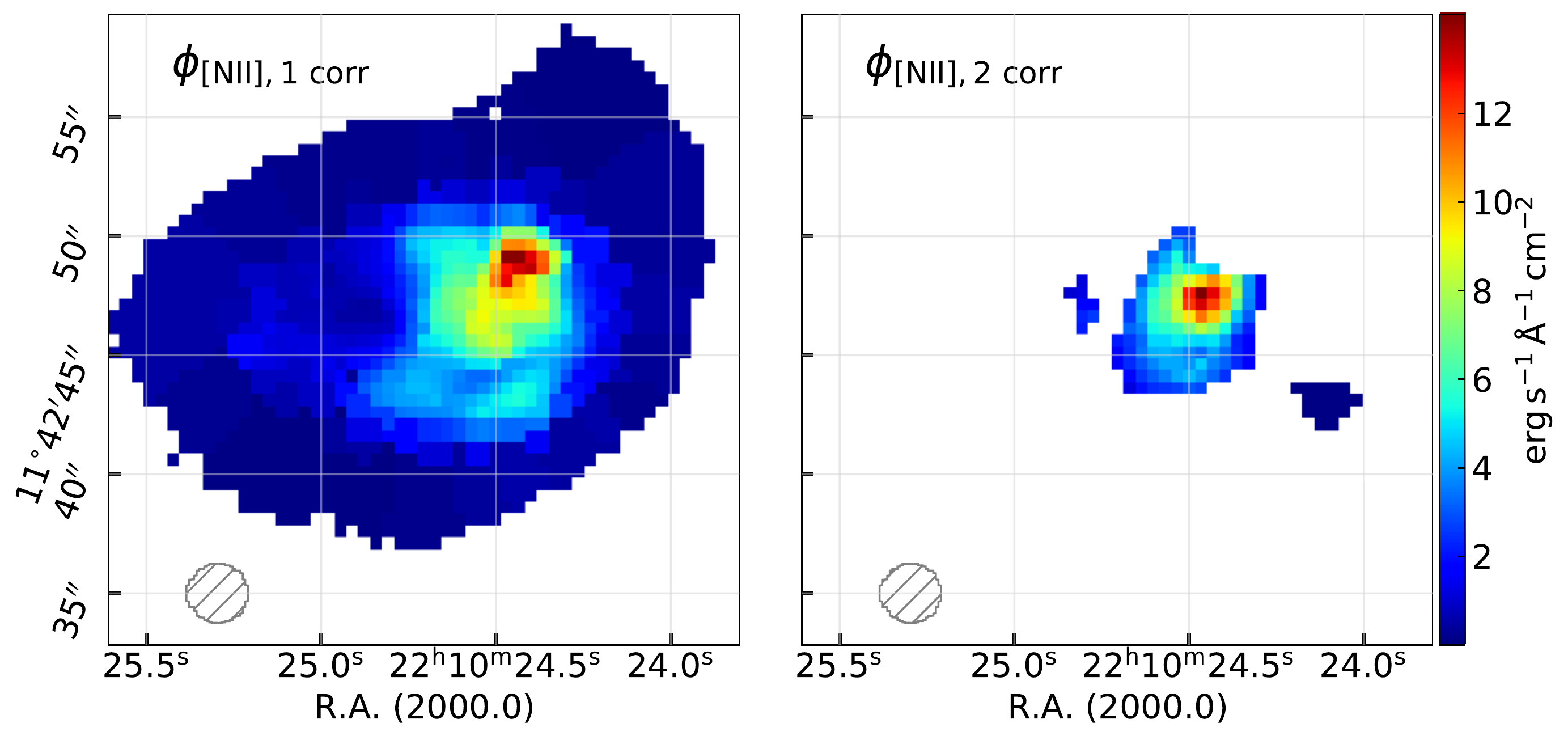}
   \caption{Respective $\mathrm{H \beta}$, [OIII]$\lambda$5007, $\mathrm{H \alpha}$, and [NII]$\lambda$6585 flux maps for both fitted components. The S/N threshold is set at 3 (spaxels with S/N below this value are masked). Moreover, for the second component, all the spaxels that do not meet the criteria defined in Sect. \ref{sect:evolution_fit} are masked. The maps shown in the right column are the respective fluxes but corrected for extinction.}
    \label{flux_maps}
\end{wrapfigure}



\clearpage
\newpage
\section{Spectra from regions with different extreme metallicities}
\begin{wrapfigure}{c}{.9\textwidth}
    \centering
    \includegraphics[width=.9\textwidth]{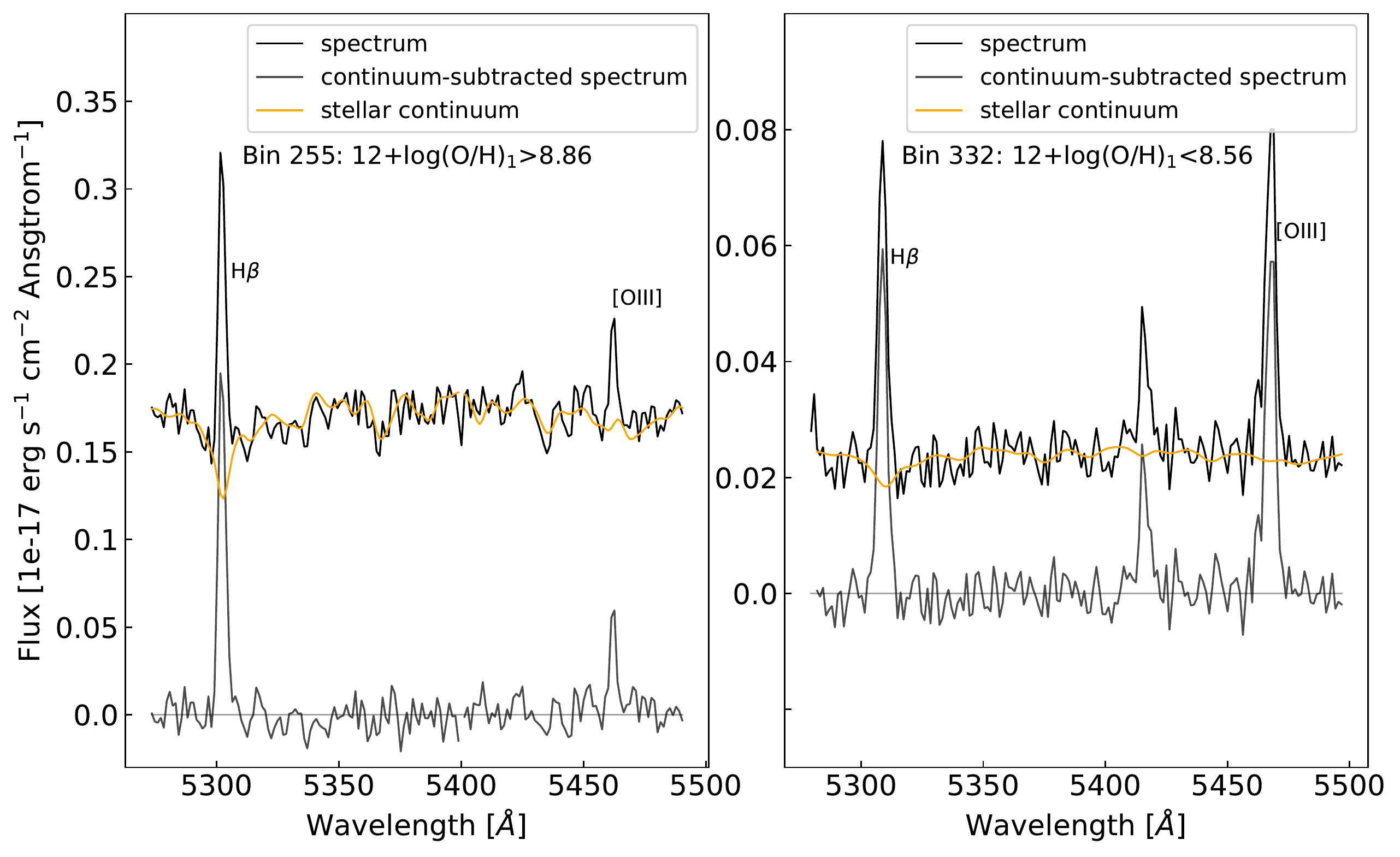}
    \caption{Different extracted spectra and their corresponding fits. The left panel shows the spectrum of a bin with a high gas-phase oxygen abundance, and the right panel shows a bin where the computed gas-phase oxygen abundance is low. In the left panel an important Balmer absorption line below $\mathrm{H \beta}$ can be seen.}
    \label{fig:spectra_with_extreme_met}
\end{wrapfigure}

\end{document}